\documentclass{aa}  
\usepackage{caption}
\usepackage{xcolor}
\usepackage[switch]{lineno}

\usepackage{graphicx}
\usepackage{txfonts}
\usepackage{multirow}
\usepackage{tabularray}
\usepackage{array}

\newcommand{\nInitialFrit}{38 } % Number of stars from Fritzewski+2025 that were fed into the modelling, some of which failed or had too large a splitting
\newcommand{\nModelledFrit}{31 } % Number of stars from Fritzewski+2025 that were successfully modelled 
\newcommand{\nModelled}{36 } % Total number of stars that were successfully modelled 
\newcommand{\nFailed}{seven } % Total number of stars for which we could not find a good model.
\newcommand{\nRadial}{24 } % Number of successfully modelled stars with at least one identified radial mode 
\newcommand{\nMultiple}{17 } % Number of successfully modelled stars with more than one identified, rotationally split non-radial mode 
\newcommand{\nDifferentialRotators}{13 } % Number of stars with more than 10% difference in their rotation rate measurements
\newcommand{\nFritzewski}{24 } % Number of stars overlapping with the 119 modelled beta Cep stars in Fritzewski et al. (2025)

\begin{document}
\newcommand{\BC}{$\beta$\,Cep }
\newcommand{\BCs}{$\beta$\,Cep stars }

   \title{Asteroseismic forward modelling of 36 $\beta$ Cep pulsators and inferences on their internal differential rotation}
   \author{M. Vanrespaille\inst{1}
          \and
          D.J. Fritzewski\inst{1}
          \and
          V. Vanlaer\inst{1}
          \and
          C. Aerts\inst{1,2,3}
          }

   \institute{Institute of Astronomy, KU Leuven,
              Celestijnenlaan 200D, 3001, Leuven, Belgium\\
              \email{mathijs.vanrespaille@kuleuven.be}
         \and
              Department of Astrophysics, IMAPP, Radboud University Nijmegen, PO Box 9010, 6500 GL Nijmegen, The Netherlands
         \and
              Max Planck Institut f\"ur Astronomie, K\"onigstuhl 17, 69117 Heidelberg, Germany         
             }
   \date{   }

  \abstract
  % context heading (optional)
   {Asteroseismic observations of the interior rotation of main sequence stars have shown that angular momentum transport is much more efficient than expected. Which transport mechanisms are responsible for this is still unclear. Detections of radial differential rotation provide valuable constraints on these transport mechanisms. Differential rotation has been detected in several massive main sequence $\beta$\,Cephei ($\beta$\,Cep) pulsators, even though fewer than ten $\beta$\,Cep stars have been asteroseismically modelled in detail so far.}
  % aims heading (mandatory)
   {We aim to expand the sample of asteroseismically forward modelled $\beta$\,Cep pulsators and exploit their potential to constrain angular momentum transport mechanisms. To that end, we seek to place constraints on their rotation profiles.}
  % methods heading (mandatory)
   {We searched for the rotational splittings of non-radial modes in a large $\beta$\,Cep sample with identified mode degrees. These were subjected to a novel forward modelling approach involving a 6-dimensional parameter space, which consistently accounts for second-order rotation effects using the state-of-the-art \texttt{StORM} oscillation code. }
  % results heading (mandatory)
   {We successfully modelled \nModelled $\beta$\,Cep stars and constrained crucial parameters such as their initial mass, internal rotation frequency, convective core mass, and age. Like in intermediate-mass stars, the internal rotation rate globally decreases in \BCs as they evolve along the main sequence. Radial differential rotation is constrained in \nMultiple \BC stars. We detect statistically significant deviations from quasi-rigid rotation in \nDifferentialRotators stars. Our constraints on eight of these \nDifferentialRotators stars indicate that the rotation rate decreases with radius between the layers probed by the identified modes while it increases in three other stars. Two stars with rotation constraints at three or more points display non-monotonic rotation profiles.}
  % conclusions heading (optional), leave it empty if necessary 
   {We affirm that radial differential rotation is common in \BC stars. Moreover, our constraints on the internal rotation are compatible with non-monotonic rotation profiles obtained from hydrodynamical simulations and rotation inversions.} 

   \keywords{Asteroseismology -- Stars: oscillations -- Stars: massive -- Stars: interiors -- Stars: evolution -- Stars: rotation}

   \titlerunning{Asteroseismic forward modelling of 36 $\beta$ Cep pulsators}
   \maketitle
   % \linenumbers
%
%-------------------------------------------------------------------

\section{Introduction}

Asteroseismology, the study of pulsating stars, has begun to unveil the interior structure of stars \citep[e.g.,][]{HekkerJCD2017, Garcia2019, Aerts2021, AertsTkachenko2024}. Thanks to the recent space photometry revolution brought on by the CoRoT \citep{Auvergne2009}, \textit{Kepler} \citep{Borucki2010}, and TESS \citep{Ricker2015} telescopes, precise asteroseismic measurements of the interior structure of thousands of stars are now available \citep[see][for an overview]{Kurtz2022}. A key goal of asteroseismology is to unravel transport processes inside the stars, notably angular momentum transport mechanisms such as those described by \citet{Zahn1992, Talon1997, Mathis2004, Mathis2005, Mathis2013, Rogers2013}. Calibrating these processes requires a measurement of the internal rotation profile of stars in various evolutionary stages \citep[e.g.,][for a review]{Deheuvels2014, Kurtz2014, Triana2015, DiMauro2016, Triana2017, Li2024, Aerts2019}.

Asteroseismic studies have constrained the rotation profiles of hundreds of intermediate-mass main sequence stars with a convective core. The vast majority of these constraints are compatible with quasi-rigid rotation regardless of the star's age \citep[][for a review]{Ouazzani2019,Li2020,Aerts2021}. These observations stand in stark contrast to the predictions of radial differential rotation increasing with age when no angular momentum is transported \citep[e.g.,][]{Maeder2009} or when modelling angular momentum transport by meridional circulation and turbulence \citep[e.g.,][]{Zahn1992, Talon1997, Decressin2009, Amard2019}. As such, angular momentum transport in intermediate-mass main sequence stars must be far more efficient than previously thought. It is currently unclear which transport mechanisms are responsible for this efficient angular momentum transport along the main sequence. As a result, current stellar evolution models are subjected to considerable systematic uncertainty. Moreover, these transport mechanisms could also transport material, further affecting the star's structure and evolution. Efficient internal mixing refuels the convective core and thus extends the star's lifetime, smoothens chemical gradients, and brings metals produced in the core to the surface. 

Our present study aims to lift our knowledge of the internal rotation properties of high-mass main sequence stars to a population level. The most massive main sequence pulsators are the $\beta$\,Cephei ($\beta$\,Cep) stars. They display low-order pressure-(p-) and gravity-(g-)modes and have masses between approximately 8 and 30 M$_\sun$ \citep{Aerts2010, Bowman2020, Kurtz2022}. While radial differential rotation is rarely detected in intermediate-mass main sequence pulsators, it has been identified in most of the few \BCs with internal rotation measurements. Such differential rotation was first reported by \citet{Aerts2003b} and was recently summarised by \citet[][]{Burssens2023}. Hence, \BCs provide valuable constraints to angular momentum transport studies in the high-mass regime. However, the exploitation of this pulsator class is limited by the small number of targets for which the internal rotation rate was measured. Fewer than ten $\beta$\,Cep stars have been asteroseismically modelled in detail and never all together in a consistent population-level study. Of these stars, only five provided constraints on their differential rotation \citep[][see their Table 1]{Burssens2023}. In three stars the rotation rate near the core is up to about 3 times faster than near the surface. In one case, the near-core region rotates slightly slower than the envelope and the final star seemingly rotates quasi-rigidly. 

CoRoT observed only one genuine $\beta$\,Cep star, which did not reveal any information on its internal rotation \citep{Degroote2009}, and \textit{Kepler} observed none with proper mode identification. This is not surprising as \BC pulsators were not the main focus of these space missions. Several hundreds of \BCs were recently discovered by TESS \citep[e.g.,][]{Shi2024,Eze2024} and other projects \citep[e.g.,][]{Burssens2019,LabadieBartz2020,Pereira2024}. However, the internal rotation profile has only been constrained in one of these stars, namely HD\,192575 \citep{Burssens2023, Vanlaer2025a}. This shortage is due to a lack of mode identifications of the detected oscillations, which are necessary to measure the interior rotation rate of $\beta$\,Cep pulsators. The internal rotation is usually measured from the rotational splittings of non-radial modes via the Ledoux constant of that splitting \citep{Ledoux1951}. As the low-order modes of $\beta$\,Cep stars do not occur in the asymptotic regimes of low or high frequencies, their Ledoux constants must be extracted from a proper stellar structure model of the star. Such a model is usually found by forward asteroseismic modelling of zonal mode frequencies, whereby the observed frequencies are matched to model predictions. Forward asteroseismic modelling is only reliable if at least a few observed modes are identified \citep{Ausseloos2004}. So far, most forward modelling efforts used a first-order perturbative treatment of rotation, which is not optimal for the \BC frequency regime \citep[][and references therein]{Suarez2009}. 

Mode identification of $\beta$\,Cep pulsations was historically performed from ground-based multi-colour photometry \citep[e.g.,][]{Heynderickx1994} or spectroscopic line profile variations \citep[e.g.,][]{Aerts2003a}. This research was boosted by long-duration multi-site observation campaigns of a few bright \BC targets \citep[e.g.,][]{Handler2004, Aerts2004b, Handler2005, Briquet2005, Handler2006, Handler2012, Briquet2012}. However, these require significant allocations of telescope time, international collaboration efforts, and intricate analyses. Consequently, such campaigns are difficult and expensive to scale up to large samples, especially for dim targets. Despite these challenges, modern ground-based multi-site monitoring projects to identify the modes in a few dozen bright $\beta$\,Cep stars are ongoing, such as the Global Asteroseismology Project \citep{Shitrit2024}. 

A new avenue towards $\beta$\,Cep mode identification recently opened up in the form of space-based multi-colour photometry. \citet{Hey2024} demonstrated that the sparse time series photometry of the \textit{Gaia} space telescope \citep{GaiaCollaboration2016, GaiaCollaboration2023} captures the dominant frequency detected by TESS in over 80\% of pulsators with an amplitude above 4\,mmag. Subsequently, \citet{Fritzewski2025} performed a multi-colour analysis on over 200 $\beta$\,Cep stars based on their amplitudes in the TESS and \textit{Gaia} passbands. For 143 of these stars, they identified the most likely degree of the dominant mode. In 33 of these stars, they also found a rotationally split multiplet which includes the dominant mode. These multiplets were subsequently matched to stellar models to estimate the envelope rotation rate. \textit{Gaia} spectroscopy placed a lower limit on the surface rotation frequency of 20 of these pulsators. Based on these two rotation frequencies, \citet{Fritzewski2025} provided upper limits on the envelope-to-surface rotation ratio of these 20 stars. These limits vary between 0.3 and 4, albeit with considerable uncertainty. 

In this study, we provide observational constraints to future theoretical studies of angular momentum transport in high-mass main sequence stars by bringing \BC asteroseismology to a population level. To that end, we revisit the sample of \citet{Fritzewski2025} looking for $\beta$\,Cep stars with sufficient mode identifications to perform forward asteroseismic modelling. Our sample of $\beta$\,Cep stars is described in Sect.\,\ref{sec:target_selection}. Section\,\ref{sec:grid_of_stellar_models} presents a new grid of stellar models and their oscillation predictions. This grid enables a novel asteroseismic modelling approach consistently including second-order rotation effects, as described in Sect.\,\ref{sec:asteroseismic_modelling_strategy}. Based on these results, Sect.\,\ref{sec:second_order_rotational_effects} evaluates the impact of the second-order rotation effects. Section\,\ref{sec:behaviour_of_rotation_rate} delves into relations between the asteroseismically inferred rotation and stellar parameters. The detections of differential rotation are shown and discussed in Sect.\,\ref{sec:differential_rotation}. Finally, a summary and our conclusions are presented in Sect.\,\ref{sec:summary_and_conclusions}.

\section{Target selection} \label{sec:target_selection}

The goal of forward asteroseismic modelling is to constrain a number of free parameters and the input physics of stellar models by matching the observed frequencies of identified modes to those predicted by those models. At minimum, this must constrain the stellar mass and age. For intermediate- and high-mass stars, core-boundary mixing \citep{Dupret2004, Mazumdar2006, Johnston2021, Pedersen2021} and envelope mixing \citep{Moravveji2015, Pedersen2021} are also essential ingredients. In \BC stars, the initial metallicity affects the modelling through degeneracies with the stellar mass and age \citep[e.g.,][]{Ausseloos2004,Dupret2004}, so it is included as a fifth free parameter. We seek to evaluate the rotation rate as well, which can be constrained from the rotational splitting of frequencies that belong to the same multiplet. 

Given the complexity of modelling in a six-dimensional parameter space, we need sufficient observational constraints. Two constraints are provided by the effective temperature and luminosity, which are now available for most known $\beta$\,Cep stars from \textit{Gaia} DR3. The remaining constraints come from observed mode frequencies, for which the azimuthal order of observed modes must be known to account for rotational shifts. Radial and zonal non-radial modes frequencies are only weakly dependent on rotation and thus reliably constrain mass, age, and internal mixing. Therefore, we constructed a sample of $\beta$\,Cep stars with at least three observed frequencies with an identified degree and azimuthal order, of which two must be part of the same multiplet to constrain rotation and two must be radial or zonal non-radial modes. No sample of $\beta$\,Cep stars with these stringent selection requirements on mode identifications has been collected or modelled before.

\subsection{Initial input from \citet{Fritzewski2025}}

Most of our sample stars are selected from the 222 \BCs presented in \citet{Fritzewski2025} and we followed their approach in drawing their stellar parameters from \textit{Gaia}. When available, the effective temperature $T_\mathrm{eff}$ comes from \textit{Gaia}'s ESP-HS pipeline \citep{Fouesneau2023}. Otherwise, we took $T_\mathrm{eff}$ from the GSPPHOT-OB pipeline. Following \citet{Fritzewski2025}, we took a conservative relative uncertainty of 10\% for $T_\mathrm{eff}$ as the errors listed in \textit{Gaia} DR3 are unrealistically small \citep{Fouesneau2023}. The luminosity $L$ was computed using \citet{Pedersen2020}'s Model 1 bolometric correction with measurements of distance, mean G magnitude, and extinction from GSPPHOT-OB. Finally, like \citet{Fritzewski2025} did, we estimated the surface rotation frequency with an unknown projection factor $f_\mathrm{rot}\,\sin{i}$ by combining the ESP-HS projected surface velocity with the radius computed from $L$ and $T_\mathrm{eff}$. 

We drew on the TESS light curves, Fourier analyses, and identifications of mode degrees of \citet{Fritzewski2025} as the foundation to build our sample on. By combining \textit{Gaia} and TESS photometry in different passbands, they assigned probabilities to the dominant frequency of each star having a degree $l$ of 0, 1, or 2. For 143 $\beta$\,Cep stars, they could identify a particular degree with a probability greater than 60\%. \citet{Daszynska2002} showed that information on the azimuthal order and inclination angle is in principle encoded in amplitude ratios of rotating \BC stars. However, the azimuthal orders of the dominant modes could not be deduced from the data-driven multi-passband identification method in \citet{Fritzewski2025}.

\subsection{Stricter mode identification from rotational splitting} \label{ssec:identifying_rotational_splitting}

To determine the degree and azimuthal order of as many significant frequencies detected in each target's TESS light curve as possible, we searched for rotationally split multiplets. For the low-order modes of single, non-magnetic $\beta$\,Cep stars, these multiplets show up as series of up to $2l+1$ roughly evenly spaced frequencies in the Fourier transform of a light curve. 

Without prior information on a star's rotation rate and the identities of its pulsation modes, there are many possible combinations of detected modes that could plausibly form a multiplet. Hence, to avoid misidentifying multiplets, we constrained ourselves to candidates that satisfy a number of conditions based on the existing observations and the expected properties of \BC pulsations. However, imposing a list of overly strict conditions can introduce selection biases and limit our sample size. Therefore, the identification of a potential multiplet was considered sufficiently secure when it satisfied at least four of these five conditions: 
\begin{enumerate}
\item All $2l+1$ frequencies in the multiplet are detected.
\item The rotational splitting in the candidate multiplet should be compatible with the $f_\mathrm{rot} \sin{i}$ estimate from \textit{Gaia} if it is available. To that end, we defined the mean rotational splitting as $\overline{\Delta f} \equiv \langle\frac{f_m - f_0}{m}\rangle$, wherein $f_m$ is the multiplet's frequency of azimuthal order $m$ (so $f_0$ is the zonal mode frequency) and $\langle \rangle$ indicates a mean over the detected non-zonal ($m \neq 0$) modes. We required that $\overline{\Delta f} > \min(f_\mathrm{rot}\sin{i} - 2\sigma_{f_\mathrm{rot}\sin{i}}, \frac{1}{2} f_\mathrm{rot}\sin{i})$ where the second term accounts for the unknown Ledoux constant as it may reduce rotational splitting by up to half in $\beta\,$Cep stars \citep{Ledoux1951}. {$\Delta f$} can be larger than $f_\mathrm{rot} \sin{i}$ due to the unknown $\sin{i}$. 
\item The identification of the candidate multiplet's degree must not contradict the multi-colour analysis of the dominant mode by \citet{Fritzewski2025}: \\
a) If the dominant mode studied by \citet{Fritzewski2025} is identified as a radial mode with a probability of at least 60\%, condition 3 was neglected since radial modes are not rotationally split. \\
b) If the dominant mode is part of the candidate multiplet, the degree suggested by the rotational splitting must have a probability of at least 40\% in \citet{Fritzewski2025}. \\
c) If the dominant mode is part of a multiplet other than the candidate multiplet, we compared $\overline{\Delta f}$ of the two multiplets. To account for the unknown Ledoux constant, we demanded $\overline{\Delta f}$ in each multiplet to differ less than a factor 2. If there are no detected frequencies in this frequency window around a high-amplitude pulsation, we took that pulsation to be a radial mode. 
\item The broader a multiplet's rotational splitting, the more likely it becomes for an unrelated mode to be misidentified as part of the multiplet. Moreover, rotational splitting in rapidly rotating stars is expected to be more asymmetric, which complicates mode identification. Therefore, we selected multiplets with $\overline{\Delta f} < 0.75 \, \mathrm{d}^{-1}$. This places an upper limit on the rotation rate of our sample, which is justified because most high-amplitude \BC pulsators occur in this rotation regime \citep[see Fig. 4 of][]{Stankov2005}. 
\item Rotationally split p-mode multiplets in slow to moderate rotators are nearly symmetric with a dimensionless asymmetry $\mathbf{A_{|m|} = -} \frac{2f_0 - f_m - f_{-m}}{f_m + f_{-m}}$ on the order of a few times 10$^{-2}$ \citep[][see their Fig.\,4 for a visualisation of $\mathbf{A_{|m|}}$]{Guo2024}. Consequently, we sought multiplets with $|A_{|m|}| < 0.10$.
\end{enumerate}

\subsection{Literature pulsators} \label{ssec:literature_stars}

To validate the novel forward asteroseismic modelling strategy described in Sect.\,\ref{sec:asteroseismic_modelling_strategy} and to expand our sample further, we added $\beta$\,Cep stars with forward modelling in the literature. Our study is focussed on pulsations of degree $l=0,1,2$ since those are the only identifications provided by \citet{Fritzewski2025} and are the most commonly observed in \BC pulsators. Due to the requirements of our modelling approach, we selected literature stars with at least one identified rotationally split multiplet and either an identified radial mode or another multiplet. 

Six well-known \BCs fulfil these conditions, namely HD\,129929 \citep[e.g.,][]{Aerts2003b, Aerts2004b, Dupret2004}, $\nu$\,Eri \citep[e.g.,][]{Aerts2004a, Ausseloos2004, DeRidder2004, Pamyatnykh2004, Jerzykiewicz2005, Dziembowski2008, Suarez2009}, $\beta$\,CMa \citep[e.g.,][]{Handler2003, Desmet2006, Mazumdar2006, Shobbrook2006}, $\theta$\,Oph \citep[e.g.,][]{Handler2005, Briquet2005, Briquet2007, Lovekin2010}, HD\,192575 \citep{Burssens2023, Vanlaer2025a, Vandersnickt2025}, and 12\,Lac \citep[e.g.,][]{Handler2006, Dziembowski2008, Desmet2009}. 12\,Lac was already included in the sample of \citet{Fritzewski2025} and their mode identification agrees with that of \citet{Handler2006} and \citet{Desmet2009}. For the other five validation stars, we adopted the observed $T_\mathrm{eff}$, $L$, frequencies, and identifications of $l$ and, if available, $m$ reported in the literature. 

\begin{figure}[t]
    \resizebox{\hsize}{!}{\includegraphics{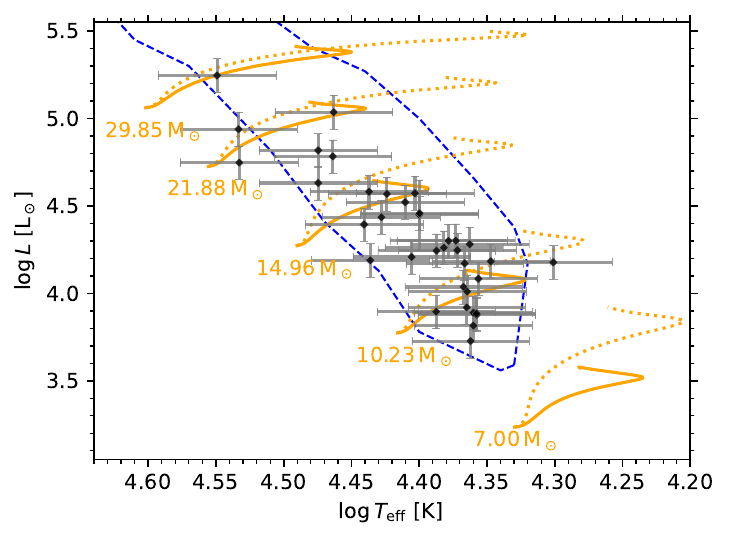}}
    \caption{Hertzsprung-Russell diagram with the targets in our sample. Evolutionary tracks from our model grid are shown in orange. Solid lines indicate the models with the weakest core-boundary and envelope mixing and dotted lines those with the strongest mixing. For reference, a p-mode instability strip of \cite{Burssens2020} is shown by a blue dashed line.}
    \label{fig:HRD_myTargets}
\end{figure}

This work presents the largest sample of $\beta$\,Cep stars with enough identified modes to perform asteroseismic forward modelling. Table\,\ref{tab:observational_input} summarises all observational input used for our modelling. Of the \nModelled stars we successfully model in Sect.\,\ref{sec:asteroseismic_modelling_strategy}, \nRadial have an identified radial mode and \nMultiple have more than one identified rotationally split multiplet. Figure\,\ref{fig:HRD_myTargets} shows these targets in the Hertzsprung-Russell diagram (HRD) alongside some evolutionary tracks from our stellar model grid.

\section{Grid of stellar models} \label{sec:grid_of_stellar_models}

Asteroseismic forward modelling requires stellar models with predictions of their pulsation frequencies to match observations to. One can either use a model grid with medium resolution or an initial grid with poor resolution which is continually refined around a star's optimal parameter values. This second approach was commonly used by past studies on individual $\beta$\,Cep stars \citep[e.g.,][]{Dupret2004, Briquet2007, Suarez2009, Burssens2023}. Since we seek to model a large number of pulsators, refining the grid for each star is not feasible. Figure\,\ref{fig:HRD_myTargets} shows our stars are spread over a broad area of the HRD, so we require a grid covering the entire $\beta$\,Cep instability strip.

\subsection{Models of stellar structure and evolution} \label{ssec:stellar_models}

We computed a new grid of non-rotating stellar models using the stellar structure and evolution code \texttt{Modules for Experiments in Stellar Astrophysics} (\texttt{MESA}) version \texttt{r24.08.1} \citep{Paxton2011,Paxton2013,Paxton2015,Paxton2018,Paxton2019,Jermyn2023}. These models were optimised for asteroseismic modelling by adopting a proper spatial resolution around burning and convective regions using the scheme of \citet{Burssens2023}. We adopt the solar metal mixture by \citet{Asplund2009} to compute solar-scaled models, relying on the helium enrichment law from Aver et al.\ (2021). We used the corresponding radiative opacity grids of the Opacity Project library \citep{Seaton2005}, except at low temperatures where they are blended with the opacity tables of \cite{Ferguson2005}. The equation of state is a blend of \texttt{FreeEOS} \citep{Irwin2004} and \texttt{Skye} \citep{Jermyn2021}. Our custom nuclear network contains 32 isotopes covering the four cold CNO-cycles and the $\alpha$-backbone up to $^{56}$Fe and $^{56}$Ni. We exclusively employed the nuclear rates of the \texttt{JINA REACLIB} \citep{Cyburt2010}. We used the Ledoux criterion for convection and set the mixing-length parameter $\alpha_\mathrm{MLT} = 2.0$. Mass loss was computed using the scheme from \cite{Vink2001}. As these mass loss estimates are known to be overestimated \citep[e.g.,][]{Puls2008,Sundqvist2011,Vink2012}, the mass loss is reduced by a factor 0.3, following \citet{Bjorklund2023}. 

We included levels of core-boundary and envelope mixing as free parameters in the models. Core-boundary mixing was implemented as exponential overshooting, which begins a distance of 0.005 pressure scale heights into the convective core. The core overshoot parameter $f_\mathrm{ov}$ varies from 0.005 to 0.035 in steps of 0.005. Beyond the overshooting zone, the models include envelope mixing caused by internal gravity waves. Based on the hydrodynamical simulations by \citet{Rogers2017}, the mixing profile is set to $D_\mathrm{mix} = D_\mathrm{mix,0} \frac{\rho_0}{\rho}$, with $D_\mathrm{mix,0}$ a free parameter, $\rho$ the local density, and $\rho_0$ the density at the base of the envelope. As shown by the simulations of \citet{Varghese2023} and confirmed by the observational calibration of \citet{Mombarg2025a}, mixing by internal gravity waves increases with mass. Therefore, we let $D_\mathrm{mix,0}$ vary from 10 to $10^6$\,cm$^2$\,s$^{-1}$ in logarithmic steps $\Delta \log{D_\mathrm{mix,0}}=1$ in the model grid. To prevent the models less massive than $13\,M_\sun$ from being completely mixed, we set the upper limit of $D_\mathrm{mix,0}$ to 10$^5$\,cm$^2$\,s$^{-1}$ in that mass regime. On the other hand, $D_\mathrm{mix,0} = 10$\,cm$^2$\,s$^{-1}$ is too low to meaningfully mix models more massive than 13\,M$_\sun$, which is required to match asteroseismic observations \citep{Moravveji2015}. As such, the lower limit for $D_\mathrm{mix,0}$ is set to 10$^2$\,cm$^2$\,s$^{-1}$ for these masses.

\begin{table}
    \centering
    \caption{Parameter space of our \texttt{MESA}-\texttt{StORM} grid. }
    \resizebox{\linewidth}{!}{
    \begin{tabular}{c c|c c c}
        \hline\hline
        parameter & code & min. & max. & number \\
        \hline
        $Z$ & \texttt{MESA} & 0.011 & 0.020 & 4 \\
        $M_\mathrm{ini}$ [M$_\sun$] & \texttt{MESA} & 7.00 & 29.85 & 43 \\
        $\log{D_\mathrm{mix,0}}$ [cm$^2$\,s$^{-1}$] & \texttt{MESA} & 1.0 ; 2.0 & 5.0 ; 6.0 & 5 \\
        $f_\mathrm{ov}$ & \texttt{MESA} & 0.005 & 0.035 & 7 \\
        $X_\mathrm{c}$ & \texttt{MESA} & 0.0001 & 0.701 & 118 \\
        $f_\mathrm{rot} / f_\mathrm{crit}$ & \texttt{StORM} & 0.0 & 0.40 & 41 \\
        \hline
    \end{tabular}
    }
    \label{tab:grid_parameters}
\end{table}

To sample the entire $\beta$\,Cep space, we computed main sequence models with 43 values for initial mass $M_\mathrm{ini}$ between 7 and 30\,M$_\sun$ in logarithmic steps of $\Delta \log{M_\mathrm{ini}} = 0.015$. This results in typical steps in $M_\mathrm{ini}$ of $0.5$\,M$_\sun$ with tighter sampling around lower $M_\mathrm{ini}$. Each \texttt{MESA} run created output for up to 118 fixed values of central hydrogen fraction $X_\mathrm{c}$ between 0.701 and 0.0001 with the resolution increasing as $X_\mathrm{c}$ decreases. This fine age resolution ensures there is a model reasonably close to the star's true parameters. The spectroscopic analyses of 31 \BC field stars by \citet{Niemczura2005} show that most \BC pulsators have a metallicity between approximately 0.011 and 0.020. The initial metallicity $Z$ covers this range in steps of 0.003. Table\,\ref{tab:grid_parameters} summarises the six-dimensional parameter space of our grid.

\subsection{Oscillation computations}

To predict the pulsation frequencies of $l=0, 1, 2$ modes in our stellar models, we employed the publicly available adiabatic oscillation code \texttt{Stellar Oscillations with Rotation} \citep[\texttt{StORM},][]{Vanlaer2026}\footnote{\url{https://storm.stellar-oscillations.org/}} version \texttt{0.1.0}, which is optimised for the low-order modes of $\beta$\,Cep stars. It includes the second-order rotation effects due to the Coriolis force and rotational deformation due to the centrifugal force. The latter is approximated by the Chandrasekhar-Milne expansion to second-order \citep{Chandrasekhar1933, Tassoul1978} applied to a spherical stellar input model. This deformation increases the star's total size and hence reduces all mode frequencies. The oscillation frequencies are further perturbed by coupling between spherical harmonics, including toroidal components \citep{Saio1981,Lee1995}. As frequencies of different azimuthal order are perturbed at different levels, these second-order rotation effects result in asymmetrical rotational splitting. 

Thanks to \texttt{StORM}'s predictions of the asymmetry of multiplets, we can include the observed rotational splitting of non-radial modes as constraints in the asteroseismic forward modelling, as also proposed by \citet{Suarez2010}. This strongly constrains $f_\mathrm{rot}$, which we include as a free parameter in our forward modelling. \citet{Mombarg2025b} compared the pulsations of 2D stellar models to those computed with \texttt{StORM} from 1D models to evaluate \texttt{StORM}'s ability to constrain an internal magnetic field from asymmetrically split multiplets. They consider rotation frequencies $f_\mathrm{rot}$ up to 20\% of the Keplerian critical rotation rate $f_\mathrm{crit}$. They showed that while the asymmetries predicted by StORM are not quite accurate enough for modelling magnetic fields in intermediate and fast rotators, they are accurate enough for asteroseismic modelling of \BC stars with identified low-order modes. 

We computed the oscillations at $f_\mathrm{rot}$ in the range 0-40\% of the Keplerian critical rotation rate $f_\mathrm{crit}$ in steps of $1\%$ of $f_\mathrm{crit}$. This pushes beyond the 0-20\% regime covered by \citet{Mombarg2025b}. As such, the modelling of some modes in the most rapidly rotating stars in our sample may be imprecise due to limitations of the rotation treatment, which is only accurate to the second order \citep[see][for a more detailed assessment of the rotation treatment]{Lee1995,Mombarg2025b,Vanlaer2026}. Indeed, while most parameters in stars with $f_\mathrm{rot} > 20\%\,f_\mathrm{crit}$ are still well-constrained, some $l=1$ multiplets are poorly reproduced, as detailed in Sect.\,\ref{sec:second_order_rotational_effects}. Nevertheless, our rotation treatment is more physically complete and thus more accurate for rotating stars than the simple first-order rotation treatments used in most past $\beta\,$Cep studies. 

Although $\beta$\,Cep stars are known to feature differential rotation, our oscillation computations assume rigid-body rotation because the rotation profiles are not known a priori. This choice limits the dimensionality and makes our modelling problem solvable while still accounting for the effects of rotation in a more advanced way than past \BC forward modelling applications. The framework outlined by \citet{Suarez2010} would permit to further constrain the rotation profiles, though this framework has not yet been applied to an observed \BC pulsator. Such an additional step is beyond our global sample study. Instead, we placed constraints on the internal rotation of our stars from uniformly rotating models in Sect.\,\ref{sec:differential_rotation}. 

We scanned for frequencies of modes with degrees $l=0, 1, 2$ between 2 and 15\,d$^{-1}$, which reliably includes the radial orders $n_\mathrm{pg}$ from -3 to +5 that we consider in our modelling. \texttt{StORM} defines $n_\mathrm{pg}$ using the Eckart-Scuflaire-Osaki scheme \citep{Eckart1961, Scuflaire1974, Osaki1975}, except for $l=1$ modes for which it uses the Takata scheme \citep{Takata2006}. The pulsation modes in deformed stars are described by a combination of spherical harmonics and thus no longer has a unique identity of $l$ and $m$ in \texttt{StORM}'s output. At the rotation rates we consider, one spherical harmonic dominates the combination though. To identify that spherical harmonic, we tracked the evolution of a deformed frequency along $f_\mathrm{rot}$ down to $f_\mathrm{rot} = 0$ where unique $l$ and $m$ are known.

\section{Asteroseismic forward modelling} \label{sec:asteroseismic_modelling_strategy}

In this section, we detail our modelling approach, demonstrate its reliability using the six validation stars in our sample, and present some essential modelling results. All observational input data used in the modelling are available in an electronic table, a snippet of which is shown in Table\,\ref{tab:observational_input}.

\subsection{Step-by-step modelling breakdown} \label{ssec:modelling_breakdown}

\begin{figure}[t]
    \centering
    \includegraphics[width=0.9\linewidth]{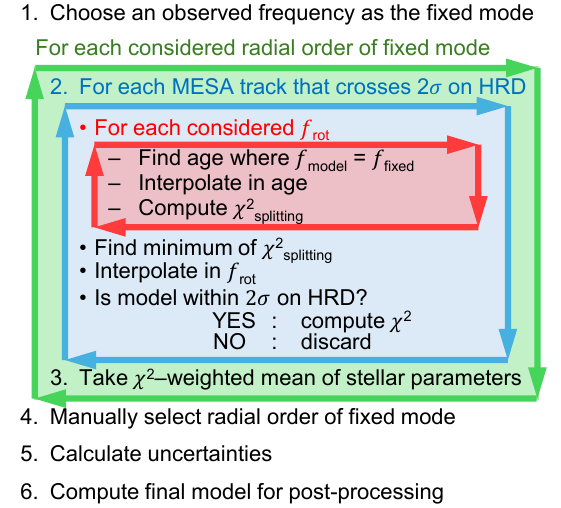}
    \caption{Step-by-step overview of the forward modelling method applied to each star. The steps in coloured boxes are repeated for each considered radial order of the fixed mode (green), for each evolutionary track in our model grid (blue), and for each considered rotation rate (red). }
    \label{fig:modelling_steps}
\end{figure}

Our forward modelling approach is inspired by that of \citet{Ausseloos2004}, who first fitted the radial mode of $\nu$\,Eri to find its age before fitting the remaining zonal modes. This two-step approach reduces the computational cost of the forward modelling, which is of great importance when modelling an entire sample of \BC pulsators. We implemented a similar methodology, but generalised it in two important ways. First, not all targets in our sample have an identified radial mode. Therefore, the age of each stellar model in the grid may be determined from a different identified mode, which we refer to as the star's `fixed mode'. Secondly, we account for the second-order effect of rotation on the mode frequencies. Consequently, the age found from the fixed mode depends on the rotation rate, so we determined the rotation rate in a consistent way during the forward modelling rather than in a `a posteriori' analysis step. 

Our modelling methodology is shown schematically in Fig.\,\ref{fig:modelling_steps}. Each of these six steps is explained in a section below.

\subsubsection{Selecting the fixed mode}

\begin{figure}
    \resizebox{\hsize}{!}{\includegraphics{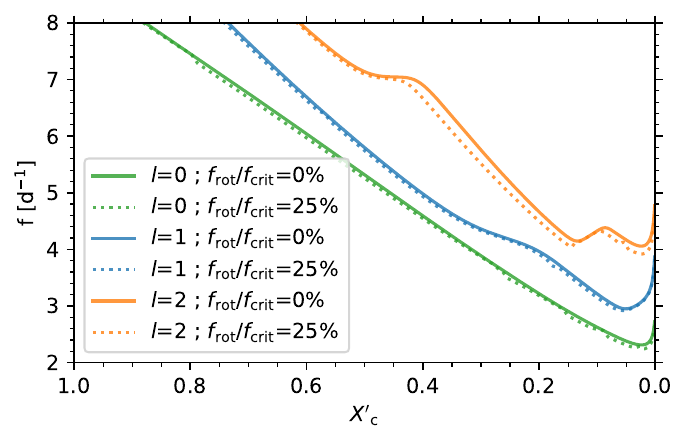}}
    \caption{Evolution of the radial (green), dipole (blue), and quadrupole (orange) zonal p$_1$-mode frequencies of a non-rotating (solid lines) and rotating (dashed lines) model. }
    \label{fig:freq_evolution_example}
\end{figure}

We first selected an identified mode as the fixed mode used to determine the optimal age. Following \citet{Ausseloos2004} and \citet{Desmet2006}, we always picked a radial mode as the fixed mode if one is identified in the observations. Figure\,\ref{fig:freq_evolution_example} shows the evolution of three zonal mode frequencies. The radial mode frequency decreases monotonically with age until the model approaches the terminal age main sequence. The frequencies of non-radial zonal modes can deviate from a linear decrease with the normalised central hydrogen mass fraction $X'_\mathrm{c}$ due to avoided crossings between p- and g-modes \citep[as discussed by][]{Aizenman1977,Burssens2023,Guo2024,Vanlaer2025a} and between modes of different degrees as a result of the rotational deformation \citep[e.g.,][]{Suarez2006,Ouazzani2012,Guo2024,Vanlaer2026}. Consequently, radial modes are preferred over non-radial modes as they allow for a unique age determination from one observed frequency in any given evolutionary track. 

For targets without an identified radial mode, we selected a zonal mode as the fixed mode since they are less dependent on $f_\mathrm{rot}$ than non-zonal modes. The deviations from linearity seen in Fig.\,\ref{fig:freq_evolution_example} are more common for modes of lower radial order and thus lower frequency. Therefore, we picked the identified zonal mode with the highest frequency as the fixed mode. Should this non-radial zonal mode be influenced by non-linear deviations, the age established from it would vary chaotically with the other free parameters. This behaviour showed up in the modelling of only one target, which was subsequently excluded from our sample as its age could not be reliably determined.

\subsubsection{Fixing age and rotation frequency for each evolutionary track} \label{sssec:age_rotation_fixing}

With a chosen fixed mode at hand and assuming its radial order is known (we revisit this assumption in Sect.\,\ref{sssec:radial_order_identification}), we compared the observed and predicted frequencies for each evolutionary track (characterised by $Z, M_\mathrm{ini}$, $\log{D_\mathrm{mix,0}}$, and $f_\mathrm{ov}$) that crosses the observed 2$\sigma$ error ellipse in the HRD. We performed cubic spline interpolation of the predicted fixed mode frequency at different ages to find the age when the model matches the observed fixed mode frequency. Subsequently, similar cubic spline interpolations were performed for all predicted frequencies and stellar parameters to produce an asteroseismic model at the fixed age. The errors incurred by this interpolation are smaller than the observational uncertainties. 

Zonal mode frequencies are changed by second-order rotation effects, as may be seen in Fig.\,\ref{fig:freq_evolution_example}. Therefore, we repeated the above age fixing for each considered rotation rate to find the relation between the fixed age and rotation rate. To select an optimal rotation frequency and thus age, we computed for each of these models at different rotation rates the merit function $\chi^2_\mathrm{splitting} = \sum_j \frac{(\Delta f_\mathrm{j,obs} - \Delta f_\mathrm{j,model})^2}{\sigma_{\Delta f_j}^2}$, with $\Delta f = |f_m - f_0|$ the rotational frequency splitting and $\sigma_{\Delta f}$ its uncertainty. The index $j$ iterates over all identified non-zonal modes. $\chi^2_\mathrm{splitting}$ represents how well a model's frequency predictions (`model') reproduce the observed rotational splitting (`obs'). Using cubic spline interpolation of $\chi^2_\mathrm{splitting}$ in $f_\mathrm{rot} / f_\mathrm{crit}$, we minimised $\chi^2_\mathrm{splitting}$ to find the best-fit value for $f_\mathrm{rot}/f_\mathrm{crit}$ in this evolutionary track. Finally, the predicted frequencies and stellar parameters in the age-fixed models are interpolated in $f_\mathrm{rot} / f_\mathrm{crit}$. 

We have thus established the age and rotation rate from the fixed mode and the measured rotational splittings. By doing so, we efficiently reduced the 118 \texttt{MESA} models along each evolutionary track with 41 sets of \texttt{StORM} mode predictions each to one optimal model and mode prediction set. We then check if that model lies within the $2\sigma$ HRD error ellipse and discard it if not.

\subsubsection{Statistical parameter estimation}

As a measure for the quality of the remaining models, we computed $\chi^2 = \sum_i \frac{(Q_\mathrm{i,obs} - Q_\mathrm{i,model})^2}{\sigma_{Q_i}^2}$. For the fitted quantities $Q_i$, we used the identified zonal mode frequencies $f_0$, identified rotational splittings $\Delta f$, $\log{T_\mathrm{eff}}$, and $\log{L}$.\footnote{Note that the frequencies of non-zonal non-radial modes are not directly fitted.} The contributions of $\log{T_\mathrm{eff}}$ and $\log{L}$ to $\chi^2$ are small due to their large observational uncertainties. As the radial orders $n_\mathrm{pg}$ of the observed modes are not known a priori, we scanned all radial orders $n_\mathrm{pg}$ in the range $(-3,\dots,5)$ and used the one with a zonal frequency closest to the observed zonal frequency of the multiplet. In the end, no observed modes were matched with the extrema $-3$ and $5$, which shows that our range in $n_\mathrm{pg}$ is sufficiently broad. 

We computed statistical parameter estimates from a weighted average over the models in the $2\sigma$ error ellipse in the HRD. The weights are given by the likelihood $\exp{(-\frac{\chi^2}{2})}$ and the local grid steps to account for the volume of the parameter each model takes up, following the method outlined in \citet{BASTA2022}. This reduces the dependence on the grid resolution by allowing the free parameters to take on values between the grid output. Moreover, the weighted standard deviation provides an estimate of the statistical uncertainty $\sigma_Q$ for each quantity $Q$.

\subsubsection{Identifying the radial orders} \label{sssec:radial_order_identification}

In the modelling steps described above, it was assumed that $n_\mathrm{pg}$ of the fixed mode was known. However, this is not the case. Therefore, we repeated steps 1 to 3  assuming different $n_\mathrm{pg}$ for the fixed mode. We then assessed the modelling outcomes to manually select the most likely $n_\mathrm{pg}$. Our selection is based on three hierarchical conditions: 
\begin{enumerate}
    \item We always favoured an assumed value for $n_\mathrm{pg}$ of the fixed mode which produced a model with frequencies  explaining unidentified significant frequencies in the observations (see green lines in Fig.\,\ref{fig:frequency_fits_validationStars}). 
    \item If the first condition did not result in a favoured $n_\mathrm{pg}$, we examined the smallest $\chi^2$ values of each modelling run and used these as a measure of the overall fit quality. We preferred a certain $n_\mathrm{pg}$ if its $\chi^2$ was less than half of those for the other $n_\mathrm{pg}$ options. 
    \item If the previous conditions did not provide a best $n_\mathrm{pg}$, we selected the model closest to the observed position in the HRD. 
\end{enumerate}

For radial fixed modes, we tested $n_\mathrm{pg}=1,2,3$. Higher order radial modes were not considered as these have never been observed in any \BC star and are predicted to be stable \citep{Pamyatnykh1999, Rehm2024}. For non-radial fixed modes, we also tried $n_\mathrm{pg}=-1,-2,-3$ on top of $n_\mathrm{pg}=1,2,3$ as \BCs are known to feature non-radial p- and g-modes. No star in our sample favoured $n_\mathrm{pg}=-3$ for its fixed mode, so we did not try higher-order g-modes. Finally, if the fixed mode has degree $l=2$, we also considered that the fixed mode may be an f-mode, represented by $n_\mathrm{pg} = 0$.

\subsubsection{Estimating uncertainties} \label{sssec:error_estimate}

The statistical uncertainty $\sigma_Q$ for each quantity $Q$ calculated in Sect.\,4.1.3 neglects the theoretical uncertainties on predicted pulsation frequencies due to uncertain model input physics. These theoretical uncertainties are typically orders of magnitude larger than the observed frequency error. Following \citet[][their Figs. 2-10]{Aerts2018}, we set the uncertainty on identified frequencies to $10^{-3}\,$d$^{-1}$ as it is a typical order of magnitude of the theoretical uncertainties for pulsation frequency predictions of $\beta\,$Cep stars.

\citet{Aerts2018} also show that the theoretical uncertainties on mode frequencies depend strongly on the mode identification, initial mass, and age. Therefore, using one value for the frequency uncertainty is a major simplification. The uncertainties should ideally be allowed to vary. To that end, we changed the likelihood distribution of each star to $\exp{(-\chi^2 / \chi^2_\mathrm{min})}$ with $\chi^2_\mathrm{min}$ the reduced $\chi^2$ of the best model for that star. This is equivalent to multiplying $\sigma_Q$ by a factor $\sqrt{\chi^2_\mathrm{min}}$. By doing so, we punish cases where $\chi^2_\mathrm{min} > 1$, meaning no grid model can reproduce the observations within the given errors. Hence, our error estimate effectively adapts to the variable theoretical frequency uncertainties. The resulting uncertainty estimates are generally conservative because $\chi^2$ is usually dominated by the rotational splitting, which mostly affects the quality of the estimate of $f_\mathrm{rot}$ rather than the other stellar parameters (cf. Sect.\,\ref{sec:second_order_rotational_effects}). 

To account for the finite resolution of our model grid, we required that the uncertainties on $Z$, $M_\mathrm{ini}$, $\log{D_\mathrm{mix,0}}$, and $f_\mathrm{ov}$ are at least half a grid step. These uncertainties on the fitted parameters are then propagated to all other stellar parameters through the correlations among the fitted parameters in the closest grid output, adopting a linear approach. This is preferable over a Monte Carlo approach, which is too demanding for the number of stars we subject to our 6-dimensional modelling procedure. Finally, as detailed in Sect.\,\ref{sec:second_order_rotational_effects}, the limitations of our rotation treatment can lead us to underestimate $f_\mathrm{rot}$ in some stars. We computed a unidirectional bias on $f_\mathrm{rot}$ to track this potential systematic error, which is taken into account in the analyses in Sects.\,\ref{ssec:specific_angular_momentum} and \ref{sec:differential_rotation}.

\subsubsection{Computing a model at the statistical parameter values}

After establishing the fixed mode's radial order, we are left with a set of statistical estimates for the six parameters ($Z$, $M_\mathrm{ini}$, $\log{D_\mathrm{mix,0}}$, $f_\mathrm{ov}$, $X_\mathrm{c}$, $f_\mathrm{rot}$). However, no \texttt{MESA-StORM} model is available in the grid at precisely these values. Hence, we computed a new \texttt{MESA-StORM} model for each star to ensure consistency in the further analyses of the results in Sects.\,\ref{sec:second_order_rotational_effects} and \ref{sec:differential_rotation}.

\subsection{Validation of the procedure} \label{sec:benchmarking}

\begin{figure*}[t]
    \centering
    \includegraphics[width=\linewidth]{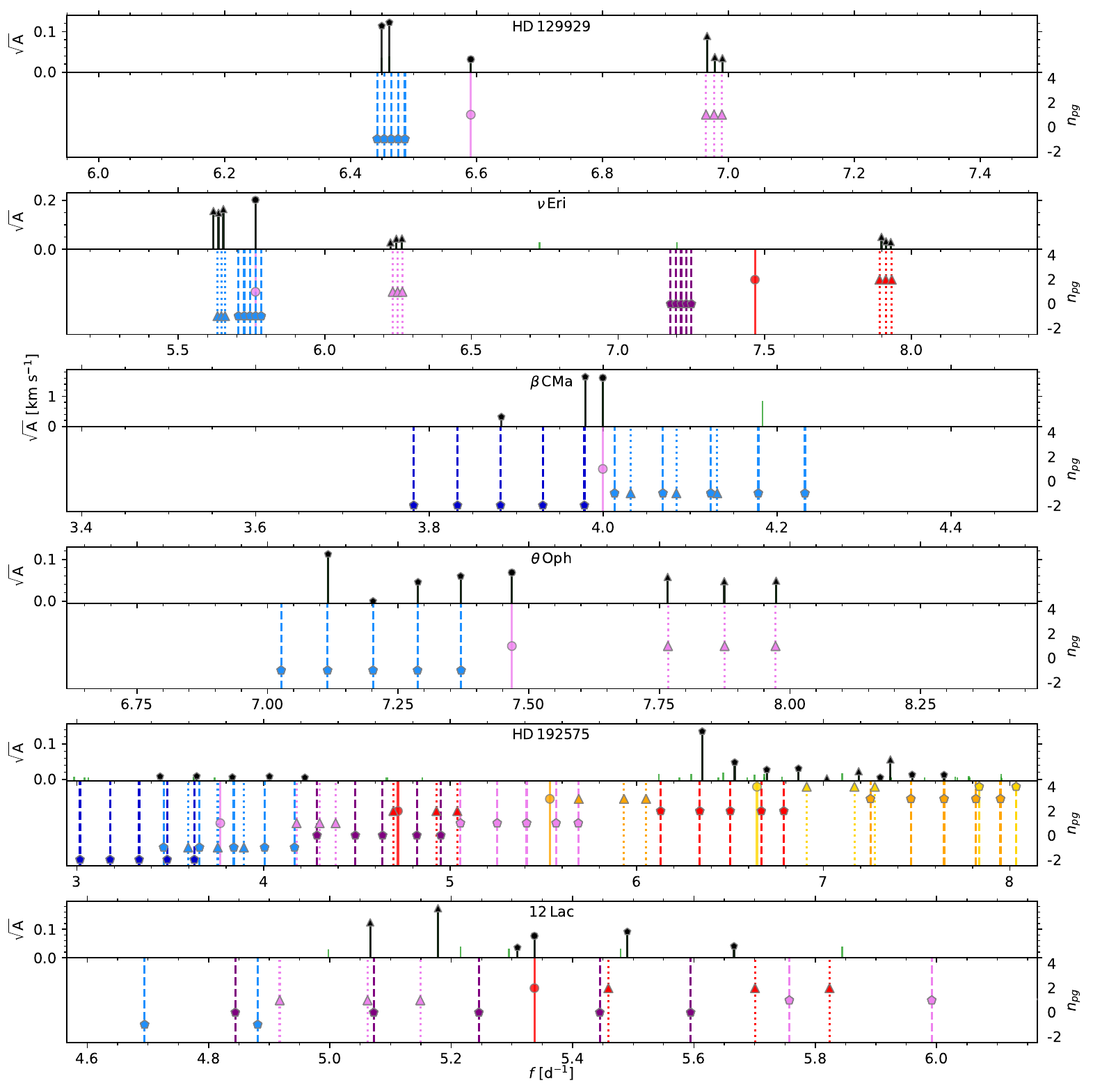}
    \caption{Comparing the observed (top) and best model's frequencies (bottom) for the six validation stars. The y-axis in the top panel shows the square root of the relative flux amplitude (except in $\beta$\,CMa, whose spectroscopically detected oscillations are given in km\,s$^{-1}$). Unidentified frequencies are green lines while identified ones are black and topped by a marker indicating the degree. In the bottom panel, the colours and y-axis indicate the radial order. }
    \label{fig:frequency_fits_validationStars}
\end{figure*}

\begin{table*}[t]
    \centering
    \caption{Comparison of the parameter estimates from our forward modelling to modelling results reported in the literature. }
    \resizebox{\linewidth}{!}{
    \begin{tabular}{m{1.6cm}|m{1.3cm}m{1.1cm}m{1.15cm}m{1.15cm}m{1.15cm}m{1.3cm}|m{0.8cm}m{.8cm}m{1.0cm}m{1.0cm}m{1.0cm}m{0.98cm}m{0.4cm}}
        \hline
        \hline
        \multicolumn{1}{c}{ } & \multicolumn{6}{c}{Our statistical model} & \multicolumn{7}{c}{Model(s) in literature} \\
        star & $Z$ & $M_\mathrm{ini}$ & $\log{T_\mathrm{eff}}$ & $\log{g}$ & $X'_\mathrm{c}$ & $f_\mathrm{rot}$ & $Z$ & $M_\mathrm{ini}$ & $\log{T_\mathrm{eff}}$ & $\log{g}$ & $X'_\mathrm{c}$ & $f_\mathrm{rot}$ & Ref. \\
         &  & [M$_\sun$] & [K] & [cm\,s$^{-2}$] &  & [d$^{-1}$] &  & [M$_\sun$] & [K] & [cm\,s$^{-2}$] &  & [d$^{-1}$] & \\
        \hline
        HD\,129929 & 0.0200(15) & 8.91(15) & 4.333(4) & 3.892(4) & 0.479(13) & 0.0132$^{+1}_{-44}$ & 0.0188 & 9.35 & 4.350 & 3.905 & 0.504 & 0.0127- 0.0147 & (1) \\
        $\nu$\,Eri & 0.0170(15) & 8.61(15) & 4.316(2) & 3.803(4) & 0.37(2) & 0.0270$^{+2}_{-2}$ & 0.0155 & 7.83 & 4.306 & 3.789 & \dots & \dots & (2) \\
        $\beta$\,CMa & 0.0133(15) & 12.6(2) & 4.380(4) & 3.653(5) & 0.23(2) & 0.060$^{+14}_{-1}$ & 0.021 & 13.5(5) & 4.373 & 3.529 & 0.184(4) & 0.054(9) & (3) \\
        $\theta$\,Oph & 0.0117(15) & 8.05(14) & 4.334(3) & 3.948(4) & 0.536(6) & 0.107$^{+1}_{-11}$ & 0.012 & 8.2(3) & 4.348(5) & 3.950(6) & 0.53(3) & 0.1068- 0.1075 & (4) \\
        HD\,192575 & 0.0200(15) & 11.8(4) & 4.366(5) & 3.598(8) & 0.33(2) & 0.186$^{+10}_{-18}$ & 0.014$^{+1}_{-1}$ & 13.0$^{+0.3}_{-1.3}$ & 4.401$^{+28}_{-5}$ & 3.66$^{+10}_{-5}$ & 0.33$^{+2}_{-15}$ & \dots & (5) \\
        \hline
        12\,Lac & 0.0136(15) & 11.8(2) & 4.367(5) & 3.674(3) & 0.225(12) & 0.265$^{+5}_{-26}$ & 0.010- 0.015 & 10.0- 14.4 & 4.343- 4.408 & 3.64- 3.70 & \dots & 0.186- 0.190 & (6) \\
        \hline
    \end{tabular}}
    \tablefoot{12\,Lac is shown separately due to its limited capacity to test our forward modelling. We compare our model of $\theta$\,Oph to that of \citet{Briquet2007} rather than the more recent study by \citet{Lovekin2010} because the latter kept the initial mass fixed. Our results for $\nu$\,Eri are compared to \citet{Ausseloos2004}'s model instead of \citet{Suarez2009} who used an initial hydrogen mass fraction of $X=0.50$ which is too different from ours to merit comparison. For consistency, the parameters of $\nu$\,Eri shown here were found using only the eight frequencies modelled in \citet{Ausseloos2004}, though the results shown throughout this work included all ten detected by \citet{Jerzykiewicz2005} and modelled by \citet{Suarez2009}. Similarly, we included 12\,Lac's $(n,l)=(0,2)$ multiplet in our modelling, which \citet{Desmet2009} reported but did not fit. We also added a new frequency detected in the TESS light curve to this multiplet. The literature results of HD\,192575 are from \citet{Vanlaer2025a} Set 1, which matches our identifications of the radial orders. }
    \tablebib{(1)~\citet{Dupret2004}; (2) \citet{Ausseloos2004}; (3) \citet{Mazumdar2006}; (4) \citet{Briquet2007}; (5) \citet{Vanlaer2025a}; (6) \citet{Desmet2009}.}
    \label{tab:benchmarking}
\end{table*}

We tested our modelling methodology on the six validation stars described in Sect.\,\ref{ssec:literature_stars}. Figure\,\ref{fig:frequency_fits_validationStars} shows that we successfully match the observed zonal frequencies except for the $l=2$ mode in 12\,Lac, which is strongly coupled to its radial mode. This makes it difficult to model that $l=2$ mode, hence why it was excluded from the observational input by \citet{Desmet2009}. The rotational splittings are reproduced very well except for some multiplets in HD\,192575 and 12\,Lac, the two most rapidly rotating stars of the six validation stars. We also find the same $n_\mathrm{pg}$ as reported in the six studies summarised in Table\,\ref{tab:benchmarking}. This shows that the conditions used for the radial order identification summarised in Sect.\,\ref{sssec:radial_order_identification} are sound.

We compare six essential stellar parameters from our modelling with the values reported in the literature in Table\,\ref{tab:benchmarking}. The uncertainties or value ranges on these parameters are included where available in the literature. Our values mostly agree with the literature within $2\sigma$. Remarkably, there is agreement in $Z$ for four validation stars despite us not including $Z$ in the observational input. There is a discrepancy in $Z$ for HD\,192575 and $\beta$\,CMa, though, which also propagates to a discrepancy in $M$ and $X'_\mathrm{c}$. In the former case, this is likely because \citet{Vanlaer2025a} considers the range $Z \in [0.012, 0.016]$, while in the latter case this is thanks to \citet{Mazumdar2006}'s observationally constrained $Z$. Overall, good agreement is obtained and differences are understood in terms of different input physics, which verifies the quality of our sample modelling.

\subsection{Summary of the modelling results} \label{ssec:summary_of_the_modelling_results}

We applied our modelling procedure to \nInitialFrit targets selected from the sample of \citet{Fritzewski2025} and found a satisfactory model for \nModelledFrit of them. Together with five stars from the literature, we have a sample of \nModelled homogeneously modelled $\beta$\,Cep pulsators. The modelling results of these \nModelled $\beta$\,Cep pulsators are summarised in Table\,\ref{tab:statistical_models}. A corner plot of the six free parameters is given in Fig.\,\ref{fig:corner_rotation_mixing_age}. Our sample covers most of the parameter space quite well, though only two stars have $M_\mathrm{ini}>20\,M_\sun$. 

The maximum value of $X'_\mathrm{c}$ is 0.64 so our sample only probes the second half of the main sequence, in line with the $\beta$\,Cep instability strips presented by \citet{Pamyatnykh1999} and \citet{Burssens2020}. A similar range in $X'_\mathrm{c}$ was reported by \citet{Fritzewski2025}, who modelled the $T_\mathrm{eff}$, $L$ and one zonal mode frequency in 119 $\beta$\,Cep stars based only on the identified $l$ of the dominant mode. The modelling outcomes of the stars in both modelled samples are compared in Appendix\,\ref{app:comparing_modelling_Fritzewski}. While there is an overall agreement in $M_\mathrm{ini}$, our modelling greatly improved the estimates of $X'_\mathrm{c}$. 

Our inability to model \nFailed stars could indicate that their mode identifications were incorrect since the identified degrees from \citet{Fritzewski2025} are not absolute. Moreover, rotational splitting can be misidentified by the chance alignment of different modes. Alternatively, the pulsation frequencies in these stars may be affected by some missing physical processes, such as binary interactions \citep[e.g.,][]{Sun2023} or magnetic fields \citep[e.g.,][]{Mathis2023, Das2024, Guo2024}. Finally, the modelling may have been affected by the limits of our framework, such as those described in the next section.

\section{Importance of second-order rotational effects} \label{sec:second_order_rotational_effects}

Thanks to the size of our sample, we can examine the systematic impact of second-order rotational effects in forward modelling of $\beta$\,Cep pulsators. Most past measurements of $f_\mathrm{rot}$ from rotational splitting $\Delta f = f_m - f_0$ in $\beta$\,Cep stars relied on first-order Ledoux splitting $\Delta f = m (1 - C_{nl}) f_\mathrm{rot}$. Therein, the Ledoux constant $C_{nl}$ was computed a posteriori from the best forward model, which was found using only the identified zonal frequencies \citep[e.g.,][]{Aerts2003b, Dupret2004, Pamyatnykh2004, Desmet2009, Burssens2023}. Even studies that included second-order rotation effects \citep[e.g.,][]{Briquet2007, Dziembowski2008, Suarez2009, Vanlaer2025a} estimated $f_\mathrm{rot}$ a posteriori. In our modelling, the rotationally induced asymmetric splitting and zonal frequency shifts play a key role in finding the best forward model (cf.\ Sect.\,\ref{sssec:age_rotation_fixing}). 

\begin{figure}[t]
    \resizebox{\hsize}{!}{\includegraphics{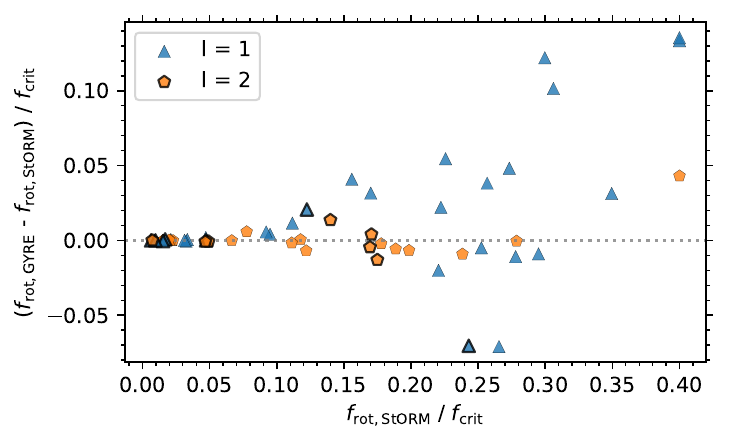}}
    \caption{Difference in rotation frequency relative to critical Keplerian rotation rate estimated from each identified multiplet by our self-consistent modelling using \texttt{StORM} and a `a posteriori' step with \texttt{GYRE} against relative rotation rate. Multiplets belonging to a validation star are outlined in black. The grey dotted line shows where the two estimates agree. }
    \label{fig:GYRE_vs_StORM_rotation_estimates}
\end{figure}

To evaluate how these second-order rotation effects in \texttt{StORM} affect the measured $f_\mathrm{rot}$, we computed $C_{nl}$ using the oscillation code \texttt{GYRE} version \texttt{7.2.1} \citep{Townsend2013}, which only includes first-order rotation effects. Emulating previous studies, we estimated $f_\mathrm{rot}$ for each rotationally split multiplet as $f_\mathrm{rot} = \overline{\Delta f} / (1 - C_{nl})$. Figure\,\ref{fig:GYRE_vs_StORM_rotation_estimates} compares these estimates, normalised by $f_\mathrm{crit}$, to our consistently optimised $f_\mathrm{rot}$ from \texttt{StORM}. Figure\,\ref{fig:GYRE_vs_StORM_rotation_estimates} shows the difference between these estimates and our consistently optimised $f_\mathrm{rot}$ from \texttt{StORM}, normalised by $f_\mathrm{crit}$ found using \texttt{StORM}. The two estimates sometimes deviate when $f_\mathrm{rot} > 10\% f_\mathrm{crit}$. We thus conclude that second-order rotational effects should not be ignored in $\beta$\,Cep modelling of stars rotating faster than approximately $10\% f_\mathrm{crit}$. 

\begin{figure}[t]
    \resizebox{\hsize}{!}{\includegraphics{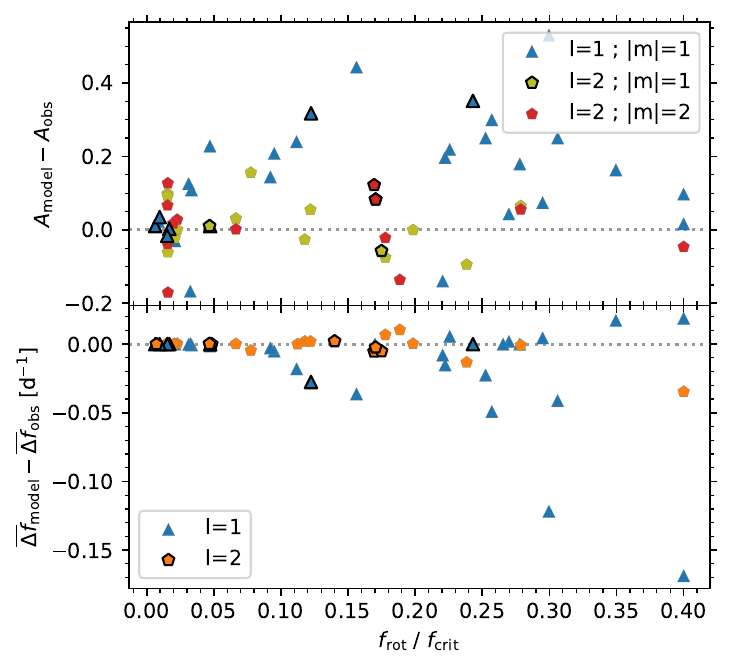}}
    \caption{Difference in asymmetry (top) and mean rotational splitting (bottom) from each multiplet between the model and observations against relative rotation rate. Multiplets belonging to a validation star are outlined in black. The grey dotted lines show where the models reproduce the observations. }
    \label{fig:discrepancies_asymmetry_meansplitting}
\end{figure}

Figure\,\ref{fig:GYRE_vs_StORM_rotation_estimates} shows that $f_\mathrm{rot}$ found with \texttt{StORM} tends to be smaller than $f_\mathrm{rot}$ estimated from \texttt{GYRE}. The difference in $f_\mathrm{rot}$ is especially prominent for $l=1$ multiplets at high $f_\mathrm{rot}/f_\mathrm{crit}$. This occurs because \texttt{StORM} often predicts highly asymmetric splittings for dipole modes, such that $\Delta f$ is greater for retrograde modes ($m<0$) than for prograde modes ($m>0$). Indeed, the top panel of Fig.\,\ref{fig:discrepancies_asymmetry_meansplitting} shows that the predicted asymmetry tends to be larger than observed for $l=1$ modes. This is partly because we looked for multiplets with a small asymmetry, as described in Sect.\,\ref{ssec:identifying_rotational_splitting}. 

Consequently, our forward modelling method reproduces the splitting of the retrograde modes reasonably well, yet sometimes underestimates the prograde mode splitting (see Fig\,\ref{fig:comparing_modelling_splittingChi2} for examples). As a result, the mean rotational splitting $\overline{\Delta f}$ can get underestimated by our modelling, as demonstrated in Fig.\,\ref{fig:discrepancies_asymmetry_meansplitting}'s bottom panel. This may imply that we underestimate $f_\mathrm{rot}$ for some stars with identified prograde $l=1$ modes rotating faster than $10\% f_\mathrm{crit}$. To quantify this effect, we computed $\overline{\Delta f}$ from the observations and our \texttt{StORM} model. A unidirectional bias on $f_\mathrm{rot}$ is then computed as $b = f_\mathrm{rot} \left(1 - \frac{\overline{\Delta f}_\mathrm{model}}{\overline{\Delta f}_\mathrm{obs}}\right)$. We only show the bias if it is greater than the uncertainty estimate of $f_\mathrm{rot}$ described in Sect\,\ref{sssec:error_estimate}. All rotational splitting data used to compute these biases and shown in Figs.\,\ref{fig:GYRE_vs_StORM_rotation_estimates} and \ref{fig:discrepancies_asymmetry_meansplitting} are gathered in Table\,\ref{tab:rotational_splitting}. 

The asteroseismic modelling methodology presented in Sect.\,\ref{ssec:modelling_breakdown} is the first applied to \BC pulsators with a consistent fit of the rotational splitting using a second-order treatment of rotation. To examine the impact of these features, we repeated our forward modelling with two alternative methods. First, we re-analysed all stars with $f_\mathrm{rot}$ optimised with a first-order rotation treatment using \texttt{GYRE} in modelling step 2. Secondly, we removed the contribution of $\Delta f$ to the $\chi^2$ cost function in step 3.The results are summarised here while details are presented in Appendix\,\ref{app:comparing_modelling_GYRE}. 

The observed zonal frequencies are better reproduced when including the second order rotation effect of stellar deformation. In this respect, the incorporation of second-order effects improves the forward modelling quality. However, the inclusion of $\Delta f$ in the $\chi^2$ cost function worsens the match with the observed zonal frequencies when modelling a star with $f_\mathrm{rot} > 20\% f_\mathrm{crit}$ using \texttt{StORM}. On the other hand, including $\Delta f$ improves the fit to the observed asymmetries and mean splittings for all $f_\mathrm{rot}$. Therefore, whether or not one should add $\Delta f$ in $\chi^2$ depends on what stellar parameters one prioritises. As this study is primarily concerned with the rotational properties of $\beta$\,Cep stars, we included $\Delta f$ in $\chi^2$. Finally, all stellar parameters other than $f_\mathrm{rot}$ are generally insensitive to these different modelling approaches. As such, the additional systematic uncertainties are modest, which justifies the exploitation of the results in Sects.\,\ref{sec:behaviour_of_rotation_rate} and \ref{sec:differential_rotation}. 

In conclusion, a first-order treatment of the rotation cannot explain multiplet asymmetries. In that sense, our sample modelling treating the rotational deformation of the stars brings an essential improvement. Nevertheless, further future improvements can be considered. Aside from the limitations in \texttt{StORM}'s treatment of the rotational deformation, the star's actual rotation profile also affects multiplet asymmetries \citep{Suarez2006,Suarez2009,Suarez2010}. Given that we assumed rigid rotation in our oscillation computations, future derivations of the rotation profiles will allow for an upgrade in the quality of the modelling of observed asymmetries. Moreover, some or all of these $\beta$\,Cep stars could possess strong internal magnetic fields as evidenced by the detection of an internal magnetic field in HD\,192575 by \citet{Vandersnickt2025}. Magnetically induced frequency shifts counteract the rotationally induced asymmetry \citep{Mathis2023,Das2024,Guo2024}.

\section{Behaviour of core mass and rotation rate} \label{sec:behaviour_of_rotation_rate}

We now look for trends in the stellar structure parameters of the modelled $\beta$\,Cep stars. In particular, we examine the mass of the convective core given its importance to the star's later evolution and chemical yields \citep[e.g.,][]{Hirschi2005,Pedersen2022,Brinkman2025}. We also focus on the total angular momentum to facilitate future studies of angular momentum transport and loss. Other correlations are shown and discussed in Appendix\,\ref{app:parameter_space}.

\subsection{Convective core mass} \label{ssec:convective_core_mass}

\begin{figure*}[t]
    \centering
    \includegraphics[width=\linewidth]{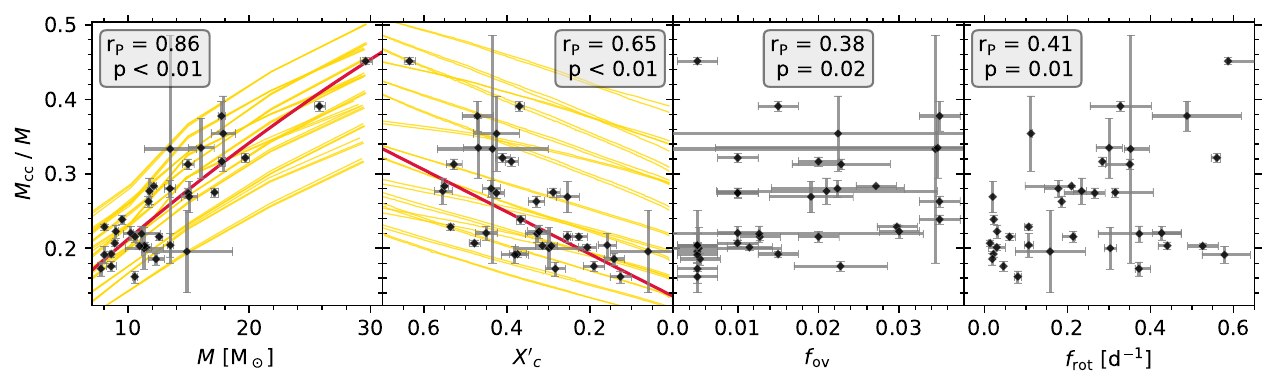}
    \caption{Convective core mass ($M_\mathrm{cc}$) relative to total current stellar mass against the current total mass ($M)$, normalised central hydrogen mass fraction ($X'_\mathrm{c})$, overshoot parameter ($f_\mathrm{ov}$), and rotation frequency ($f_\mathrm{rot}$). Red lines show a quadratic and linear fit of $M_\mathrm{cc}$ against $M$ and $X'_\mathrm{c}$, respectively. Thin gold lines show some relations from our model grid between $M_\mathrm{cc}$ and $M$ and $X'_\mathrm{c}$ with the other input parameters kept fixed. Boxes display the Pearson correlation coefficients and the p-values of the associated t-tests. } 
    \label{fig:relative_core_mass_relations}
\end{figure*}

The convective cores of massive stars shrink as they evolve along the main sequence due to the decreasing opacity as hydrogen is fused into helium. This has already been studied from asteroseismology of $\gamma\,$Dor \citep[e.g.,][]{Mombarg2021} and SPB stars \citep[e.g.,][]{Pedersen2022}. With the forward modelling results of our $\beta$\,Cep sample, we can now asteroseismically calibrate the behaviour of the relative convective core mass $M_\mathrm{cc}/M$ across a higher mass range. To that end, we fitted $M_\mathrm{cc}/M$ against the current mass $M$ and $X'_\mathrm{c}$, as shown in Fig.\,\ref{fig:relative_core_mass_relations}. The best fitting relations are given by
\begin{align*} 
    M_\mathrm{cc} / M &= -9(18)\,10^{-5}\,\left( \frac{M}{\mathrm{M_\sun}}\right)^2 + 0.0155(6)\,\frac{M}{\mathrm{M_\sun}} + 0.07(5) \\ 
    M_\mathrm{cc} / M &= 0.28(6)\,X'_\mathrm{c} + 0.14(3) \\ 
    M_\mathrm{cc} / M &= -1.9(6)\,10^{-4}\,\left( \frac{M}{\mathrm{M_\sun}}\right)^2 + 0.017(2)\,\frac{M}{\mathrm{M_\sun}} \\
    & \hspace{1cm} + 0.216(14) \, X'_\mathrm{c} - 0.012(18)
\end{align*}
The coefficients of determination of these three relations are 0.73, 0.34, and 0.92, respectively, which indicates that the bivariate fit explains most of the variability in $M_\mathrm{cc}/M$. We note that these relations are model dependent, as evidenced by the model relations shown in gold in Fig.\,\ref{fig:relative_core_mass_relations}. Our results are in agreement with \citet{Johnston2021}, who compared $M_\mathrm{cc}$ from asteroseismology and eclipsing binaries to conclude that stars with a convective core possess a wide range of core masses. 

The rightmost panel of Fig.\,\ref{fig:relative_core_mass_relations} shows the correlation between $M_\mathrm{cc}/M$ and $f_\mathrm{rot}$, which is due to the $f_\mathrm{rot} - X'_\mathrm{c}$ correlation seen in Fig.\,\ref{fig:corner_rotation_mixing_age}. This relation between $f_\mathrm{rot}$ and $X'_\mathrm{c}$ reflects the decrease of the internal rotation rate along the main sequence. Such a decrease is in line with the large sample of intermediate-mass main sequence stars shown in \citet{Aerts2021}'s Fig.\,6 and \citet{Aerts2025a}'s Fig.\,9. Another significant correlation included in Fig.\,\ref{fig:corner_rotation_mixing_age} is between $f_\mathrm{ov}$ and $X'_\mathrm{c}$, which indicates that core overshooting weakens as the stars evolve. This enhances the shrinking of the core size with age because overshooting provides the core access to more hydrogen from the envelope and hence increase core size. Parametrised core overshooting is merely a reflection of core boundary mixing due to a multitude of instabilities occurring in this transition layer. Many of these instabilities are caused by the local rotation rate \citep[see][for extensive discussions of these instabilities]{Heger2000,Aerts2019}, meaning these last two correlations are likely connected.

\subsection{Specific angular momentum of $\beta$\,Cep stars} \label{ssec:specific_angular_momentum}

\begin{figure*}[th]
    \sidecaption
    \includegraphics[width=12cm]{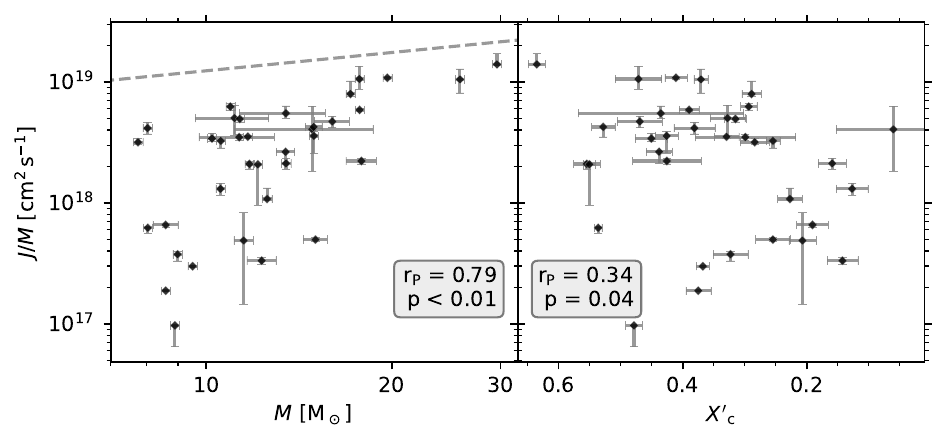}
    \caption{Specific angular momentum of each star against its mass (left) and core hydrogen mass fraction (right). On the left panel, the grey dashed line indicates  the upper limit on the specific angular momentum for stars more massive than $2.5\,M_\sun$ derived by \citet{Aerts2025b}. Boxes show the Pearson correlation coefficients and the p-values of the associated t-tests.} 
    \label{fig:JM_against_M_Xc}
\end{figure*}

The dependence of the specific angular momentum $J/M$ of a population of stars on mass and age has long been used to study initial stellar rotation rates and angular momentum losses \citep[e.g.,][]{Kraft1967,Kawaler1987,Kawaler1988}. Assuming rigid-body rotation, it is computed as $J/M = 4\pi/3\, f_\mathrm{rot} R^2$. Recently, \citet{Aerts2025b} estimated $J/M$ of approximately 3000 main sequence pulsators with masses $M \in [1.3, 9]\,\mathrm{M}_\sun$ from the near-core rotation rate and assuming quasi-rigid rotation. Figure\,\ref{fig:JM_against_M_Xc} shows our \BC stars in the $M \in [7, 30]\,\mathrm{M}_\sun$ range lie below the upper limit of \citet{Aerts2025b}. This was expected since more massive stars lose more mass and angular momentum from stellar winds. Their $J/M$ upper limit is over 50\% greater than the $J/M$ of our stars, which is in line with the level of differential rotation observed in $\beta$\,Cep stars \citep[][ cf.\ Sect.\,\ref{sec:differential_rotation}]{Burssens2023}. As such, while our $J/M$ may be inaccurate due to differential rotation, our stars still obey the upper limit. 

\citet{Aerts2025b} found evidence of differential rotation for the B-type stars in their sample. In contrast to \citet{Aerts2025b} however, there is no sign of $J/M$ increasing as $X'_\mathrm{c}$ decreases in the right panel of Fig.\,\ref{fig:JM_against_M_Xc}. The lack of such a relation may indicate that the internal rotation profiles of our massive stars differ from those of the intermediate-mass stars. We investigate this in the next section.

\section{Differential rotation constraints in \nMultiple $\beta$\,Cep stars} \label{sec:differential_rotation}

\begin{figure}[t]
    \resizebox{\hsize}{!}{\includegraphics{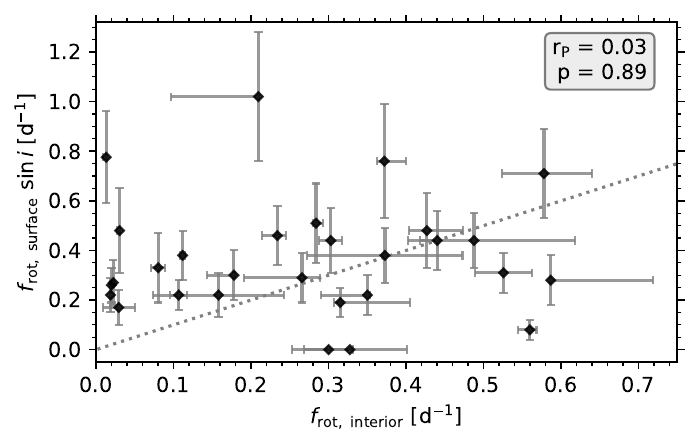}}
    \caption{Projected surface rotation frequency from \textit{Gaia}'s ESP-HS pipeline against the interior rotation frequency when modelling all identified multiplets. The grey dotted line marks where the two measurements are equal. The Pearson correlation coefficient and the p-value of the associated t-test is included in a box. Two stars were assigned a surface rotation velocity of zero, which indicates that rotational broadening was not detected. }
    \label{fig:interior_to_surface_rotation}
\end{figure}

We now seek to constrain the rotation profiles in a subsample of our $\beta$\,Cep stars. Following \citet{Fritzewski2025}, we placed an upper bound on the envelope-to-surface rotation ratio from our optimised internal rotation frequencies $f_\mathrm{rot,interior}$\footnote{The asteroseismically estimated internal rotation rate was previously symbolised by $f_\mathrm{rot}$. We add the `interior' subscript here to distinguish it from the surface rotation rate estimated from \textit{Gaia} DR3.} and the projected surface velocities $f_\mathrm{rot,surface}\sin{i}$ estimated from \textit{Gaia} spectroscopy. Figure\,\ref{fig:interior_to_surface_rotation} shows the relation between these two rotation measurements for 29 stars with an available estimate of $f_\mathrm{rot,surface}\sin{i}$. Notably, they are uncorrelated, which is in part because $f_\mathrm{rot,surface}\sin{i}$ is only a lower limit of $f_\mathrm{rot,surface}$. On the other hand, broadening of spectroscopic lines by the pulsations can lead to $f_\mathrm{rot,surface}\sin{i}$ getting overestimated for our high-amplitude \BC pulsators. Therefore, we turn to purely asteroseismic constraints on the internal rotation profile, which are more precise and unaffected by unknown projection factors.

\begin{figure*}[t]
    \centering
    \includegraphics[width=\linewidth]{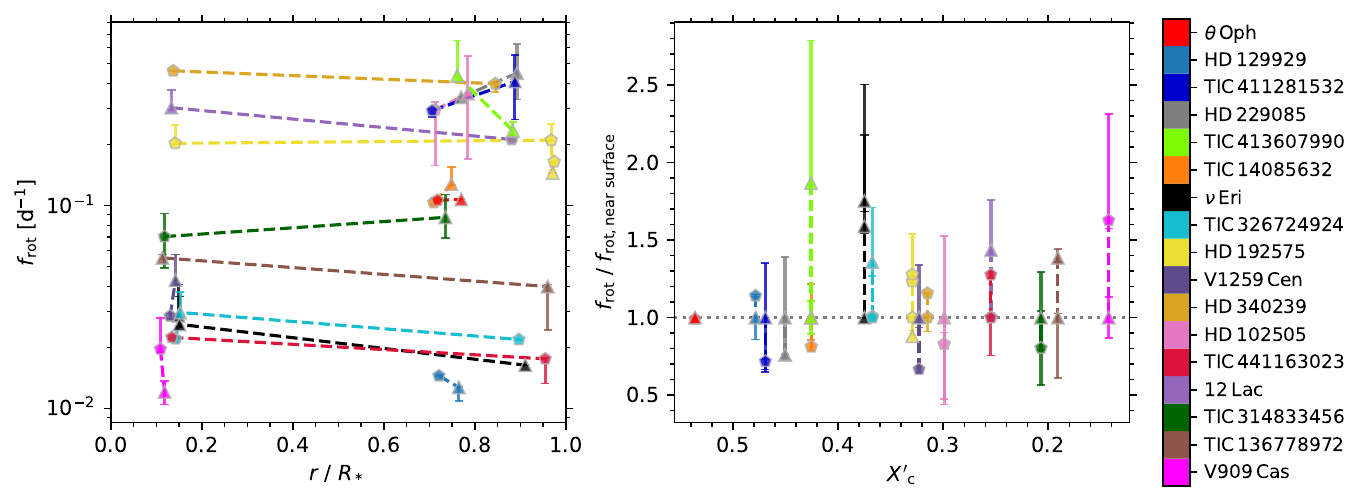}
    \caption{Interior differential rotation in \nMultiple stars with several rotationally split multiplets. The left panel shows the rotation frequency from a rotationally split multiplet at the radius where that multiplet is most sensitive. Rotation frequencies extracted from $l=1$ and $l=2$ multiplets are marked as triangles and pentagons, respectively. The right panel shows these rotation rates normalised by the rotation rate nearest the surface against the core hydrogen mass fraction. Each star has a unique colour indicated in the colourbar, which is sorted by normalised central hydrogen mass fraction. }
    \label{fig:diffrot_log_atMode_against_Xc}
\end{figure*}

In order to constrain the internal differential rotation of stars using rigidly rotating models, we re-analysed the \nMultiple stars with more than one rotationally split multiplet once per multiplet. Each time, we included all identified zonal mode frequencies in the observational input, but only one multiplet's rotational splitting $\Delta f$. Consequently, $f_\mathrm{rot}$ is optimised for only that multiplet in modelling step 2. For consistency, the estimate of $f_\mathrm{rot}$ in step 3 used the same model weights found when modelling all multiplets' $\Delta f$. Next, we computed the multiplet's rotational sensitivity kernel $K_{nl}$ from the final model computed in modelling step 6. Figure\,\ref{fig:kernels_examples} shows some examples of sensitivity kernels and Appendix\,\ref{app:sensitivity_kernels} further discusses these kernels and their sensitivity to the model parameters. As a crude constraint on the rotation profiles, the left panel of Fig.\,\ref{fig:diffrot_log_atMode_against_Xc} shows $f_\mathrm{rot}$ for each multiplet against the normalised radius where it is most sensitive. The right panel plots the rotation rate normalised to that of the mode that is most sensitive closest to the surface. Normalised rotation rates below 1 thus suggest that the rotation rate increases with radius. All data presented in this section are collected in Table\,\ref{tab:differential_rotation}. 

For each combination of two $f_\mathrm{rot}$ measurements in a star, we performed the statistical test described in Appendix\,\ref{app:statistical_test}, which considers both the modelled uncertainty on each $f_\mathrm{rot}$ and the unidirectional bias described in Sect.\,\ref{sec:second_order_rotational_effects}. In \nDifferentialRotators of the \nMultiple stars, $f_\mathrm{rot}$ varies significantly, which indicates these stars display radial differential rotation. The levels of radial differential rotation agree with the constraints on differential rotation of \BCs reported in Table\,1 of \cite{Burssens2023} and Fig.\,17 of \citet{Fritzewski2025}. Similar levels have recently been inferred for intermediate-mass g-mode pulsators as well \citep{Aerts2025a}. We find that $f_\mathrm{rot}$ most commonly decreases with radius $r$ as observed before \citep{Burssens2023}. For four stars $f_\mathrm{rot}$ increases with $r$, which has not been detected in $\beta$\,Cep pulsators before. \citet{Briquet2007} suggested that $\theta$\,Oph's rotation profile might slightly increase outwards, although they found it to be consistent with a flat rotation profile, which we corroborate. 

Differential rotation is detected in \nDifferentialRotators out of \nMultiple stars despite some biases against it. First, the low-order modes in $\beta$\,Cep stars are sensitive to broad regions of the stellar envelope, as shown in Fig.\,\ref{fig:kernels_examples}. When the sensitivity kernels of two multiplets overlap, the difference in the measured $f_\mathrm{rot}$ is diminished as $f_\mathrm{rot}$ is averaged over similar layers. Consequently, even the four stars without a significant difference in $f_\mathrm{rot}$ might be differential rotators too. Secondly, there is a selection bias in our mode identification from rotational splitting. As discussed in Sect.\,\ref{ssec:identifying_rotational_splitting}, candidate multiplets with rotational splitting differing by more than a factor 2 were excluded from our sample. Finally, our modelling is set up to reproduce all observed rotational splittings based on models assuming solid-body rotation. Consequently, the difference in $f_\mathrm{rot}$ required to optimally reproduce the splitting of each multiplet individually is reduced. Given that we detect differential rotation in so many stars despite these biases, we conclude that radial differential rotation is common in \BC stars.

Remarkably, $f_\mathrm{rot}$ appears to increase with $r$ in three of the five stars with two multiplets probing $r/R_*>0.5$, namely HD\,229085 (grey in Fig.\,\ref{fig:diffrot_log_atMode_against_Xc}), $\theta$\,Oph (red), and TIC\,14085632 (orange). We emphasise that the rotation values in Fig.\,\ref{fig:diffrot_log_atMode_against_Xc} result from broad sensitivity kernels that partially overlap and that the difference in $f_\mathrm{rot}$ is comparable to the errorbars in these three stars. Nonetheless, such upwards rotation profiles can be physically explained by strong internal angular momentum transport or accretion from a binary companion, which massive main sequence stars commonly have \citep{Sana2012,Offner2023}. In six out of eight stars with one multiplet most sensitive near the core and another in the upper envelope, $f_\mathrm{rot}$ decreases with $r$. Combined, these trends indicate that $\beta$\,Cep pulsators might have non-monotonic rotation profiles, featuring a relatively rapidly rotating core and a slower envelope with a local outward increase in $f_\mathrm{rot}$. Our rotation constraints in two of the three stars with three or more rotationally split multiplets, HD\,192575 (yellow) and TIC\,326724924 (cyan), also suggest a non-monotonic rotation profile. The non-monotonicity of HD\,192575's rotation profile is further confirmed by the tests on the rotational splitting suggested by \citet{Reese2015}, as detailed in Appendix\,\ref{app:Reese_test}. The rotation inversions of HD\,192575 by \citet{Vanlaer2025a} show that it has a core-to-envelope rotation ratio no greater than two and its rotation profile is not monotonic, which we confirm. Furthermore, both 2D hydrodynamical simulations \citep[e.g.,][]{Rogers2025} and 1D stellar structure models including angular momentum transport by waves \citep[e.g.,][]{Neiner2020} can produce such non-monotonic rotation profiles in high-mass main sequence stars.

\section{Summary and conclusions} \label{sec:summary_and_conclusions} 

We presented a sample of \nModelled asteroseismically modelled $\beta$\,Cep stars with the aim to exploit their potential to constrain angular momentum transport on the main sequence. This marks the first forward asteroseismic modelling of a population of $\beta$\,Cep stars. We developed and applied a novel forward modelling methodology to the sample in a homogeneous way. Our modelling is based on a new grid of \texttt{MESA} main sequence models with a wide range in initial mass, metallicity, age, core overshoot, and envelope mixing. The oscillations of these models were computed across a broad range of rotation rates using the new public \texttt{StORM} code, which takes into account centrifugal deformation and second-order rotational effects. Our grid of stellar models and their oscillations are publicly available (see Data Availability section).

Our modelling procedure constrained key stellar parameters such as the mass, age, convective core mass, mixing of chemicals, and internal rotation rate. We exploited our sample to calibrate the mass- and age-dependency of the convective core mass. We found that core overshooting weakens with age. Consequently, stellar models should include evolving core mass fractions, e.g. by using a time-dependent core overshooting scheme. 

Like in intermediate-mass main sequence stars, the global rotation rate decreases as the stars evolve. Moreover, $\beta$\,Cep stars obey the specific angular momentum relations of intermediate-mass stars. We constrained radial differential rotation in \nMultiple stars, increasing the sample of $\beta$\,Cep stars with such asteroseismic measurements by more than a factor three. An overall trend emerged that the inner region near the convective core usually rotates faster than the envelope. We found that that the rotation rate may increase with radius in the outer envelope. Such a rotation gradient is expected for stars that experienced accretion in their recent past or are subject to the action of internal gravity waves. We also conclude that non-monotonic differential rotation occurs in at least some $\beta$\,Cep stars. Testing which angular momentum transport mechanisms can explain these rotation profiles will require stellar modelling using new generations of stellar structure and evolution models with realistic transport mechanisms in order to verify if they reproduce all the observed rotational splitting. 

While our modelling framework includes rotational effects in a more consistent manner than any previous forward modelling of \BC stars, it revealed some shortcomings. An example is the overestimation of the asymmetry of rotationally split dipole multiplets, which indicates some important physics is missing or oversimplified in our modelling. Internal magnetic fields are an obvious candidate as they are known to reduce the asymmetry of rotationally split multiplets. However, unambiguously disentangling the effects of both magnetism and rotation from asymmetries alone is difficult. This issue is alleviated if a multiplet displays both rotational and magnetic splitting, resulting in more than $2l+1$ components. HD\,192575 is currently the only known $\beta$\,Cep star with such splitting \citep{Vandersnickt2025}. Consequently, a population level study including magnetic effects is not plausible until more $\beta$\,Cep pulsators with magnetic splitting are discovered. 

An observational avenue to continue improving our understanding of $\beta$\,Cep stars is to further expand the sample to fill out the broad parameter space. In particular, there is still a shortage of $\beta$\,Cep stars more massive than 20\,M$_\sun$. Further, more targets with enough rotationally split multiplets to measure the internal rotation rate near the core and throughout the envelope at different ages would also be highly beneficial. The upcoming PLATO space telescope \citep{Rauer2025} is capable of providing the necessary photometry to detect the pulsation frequencies of a few tens of $\beta$\,Cep pulsators with high precision \citep{Nascimbeni2025}. These frequencies may then be identified by ground- or space-based multi-colour photometry and from rotational splitting.

\section*{Data availability}

The complete versions of Tables \ref{tab:observational_input}, \ref{tab:statistical_models}, \ref{tab:rotational_splitting}, and \ref{tab:differential_rotation} are available in electronic form at \url{https://github.com/Mathijs-Vanrespaille/Vanrespaille_BetaCepheiForwardModelling.git} (placeholder, to be replaced with CDS information). The \texttt{Python} code used to perform our forward modelling is publicly available at \url{https://github.com/Mathijs-Vanrespaille/AMPHAROS}. The \texttt{MESA} stellar models, including the \texttt{StORM} oscillation computations, can be freely accessed at the KU Leuven Research Data Repository (link to be added). Meanwhile, the \texttt{MESA} work directory and \texttt{StORM} setup can be found at the following Zenodo repository (link to be added). Electronic figures similar to Figs.\,\ref{fig:frequency_fits_validationStars} and \ref{fig:kernels_examples} for all \nModelled $\beta$\,Cep stars with a good model in our sample are available at the Zenodo repository (link to be added).

\begin{acknowledgements}

The authors thank the two anonymous referees for their helpful feedback. MV warmly thanks Hannah Brinkman for her advice in updating the \texttt{MESA} inlists and for designing the nuclear network and acknowledges Ehsan Moravveji for his assistance in setting up the \texttt{MESA} grid computations. He also thanks Noi Shitrit, Tami Rogers, Miriam Rodriguez-Sanchez, Jelle Vandersnickt, Zhao Guo, and Pablo Huijse for useful conversations. Helpful comments on the manuscript were provided by Alex Kemp, Joey Mombarg, Michel Rieutord, Laura Scott, and Keegan Thomson-Paressant. The \texttt{MESA} and \texttt{StORM} computations were done at the Flemish Supercomputer Centre (VSC). MV has received funding from the KU Leuven Research Council (doctoral mandate grant DB/24/008). VV gratefully acknowledges support from the Research Foundation Flanders (FWO) under grant agreement N$^\circ $1156923N (PhD Fellowship). DJF and CA acknowledge support from the Flemish Government under the long-term structural Methusalem funding program by means of the project SOUL: Stellar evolution in full glory, grant METH/24/012 at KU Leuven. MV and CA acknowledge financial support from the European Research Council (ERC) under the Horizon Europe programme (Synergy Grant agreement N$^{\circ}$101071505: 4D-STAR). While partially funded by the European Union, views and opinions expressed are however those of the authors only and do not necessarily reflect those of the European Union or the European Research Council. Neither the European Union nor the granting authority can be held responsible for them. CA also acknowledges the Belgian Federal Science Policy Office (BELSPO) for the provision of financial support in the framework of the PRODEX Programme of the European Space Agency (ESA). 
\newline
\textbf{Software:} This research made use of the \texttt{astropy} \citep{AstropyCollaboration2022}, \texttt{h5py} \citep{Collette2013}, \texttt{matplotlib} \citep{Hunter2007}, \texttt{NumPy} \citep{Harris2020}, \texttt{pandas} \citep{McKinney2010}, and \texttt{SciPy} \citep{SciPy2020} \texttt{Python} packages. We used the \texttt{MESA} \citep{Paxton2011, Paxton2013, Paxton2015, Paxton2018, Paxton2019, Jermyn2023} stellar structure and evolution code to compute our new grid of stellar models and the codes \texttt{StORM} \citep{Vanlaer2026} and \texttt{GYRE} \citep{Townsend2013} to perform the oscillation computations.

\end{acknowledgements}

\bibliographystyle{aa}
\bibliography{my_bib}

\begin{appendix}

\section{Parameter space and correlations} \label{app:parameter_space}

\begin{figure*}[t]
    \centering
    \includegraphics[width=\linewidth]{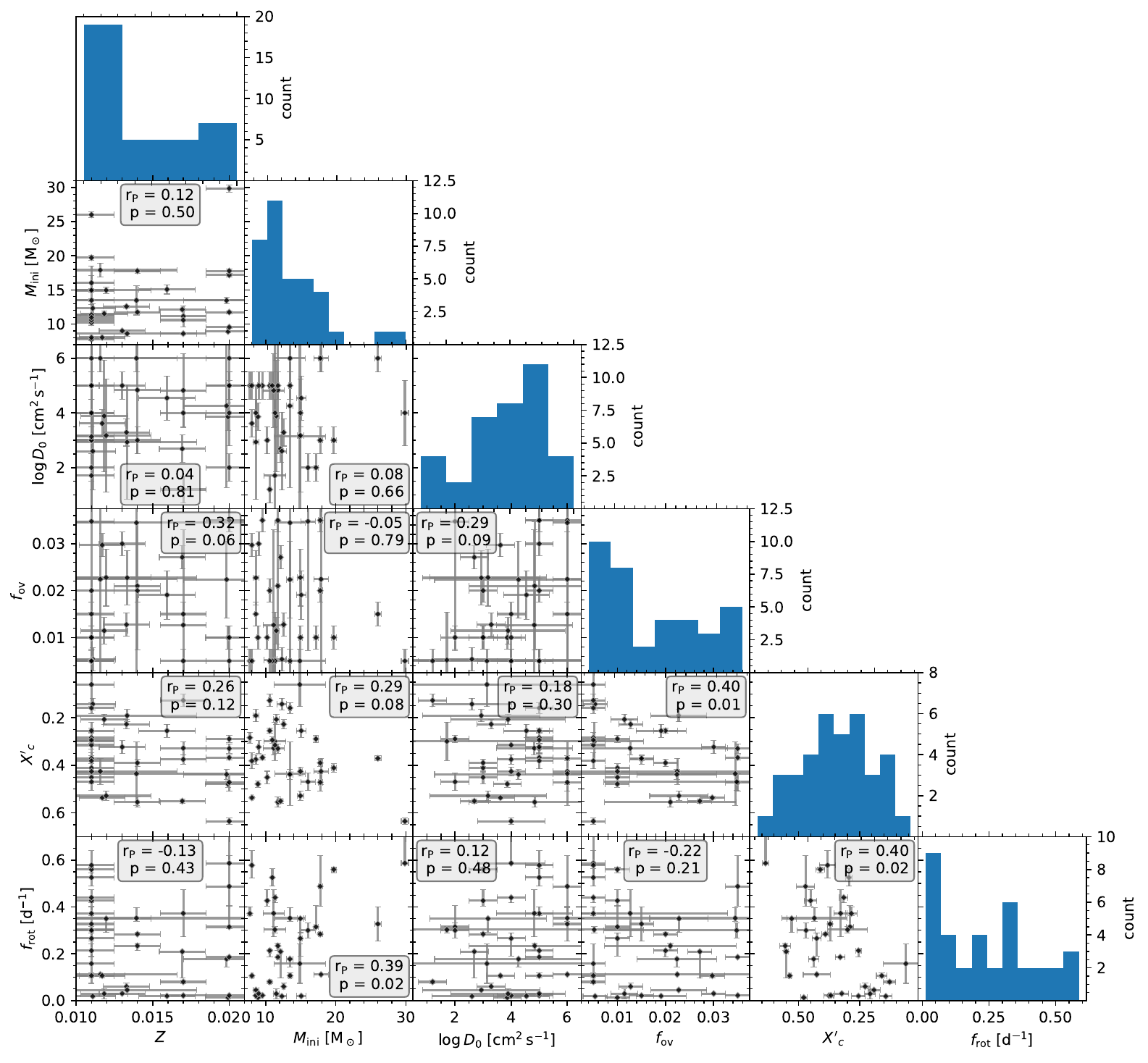}
    \caption{Corner plot of the initial metallicity, initial mass, envelope mixing strength, core overshoot parameter, normalised central hydrogen mass fraction and rotation frequency of the 36 successfully modelled stars. Pearson correlation coefficients and p-values of the associated t-tests between each set of two parameters are included in boxes. }
    \label{fig:corner_rotation_mixing_age}
\end{figure*}

Figure\,\ref{fig:corner_rotation_mixing_age} shows a corner plot of the six free parameters in our modelling. Some of these distributions and correlations were already discussed in Sects.\,\ref{ssec:summary_of_the_modelling_results} and \ref{ssec:convective_core_mass}, respectively. Here we examine some of the remaining correlations or lack thereof. 

\begin{figure}[t]
    \resizebox{\hsize}{!}{\includegraphics{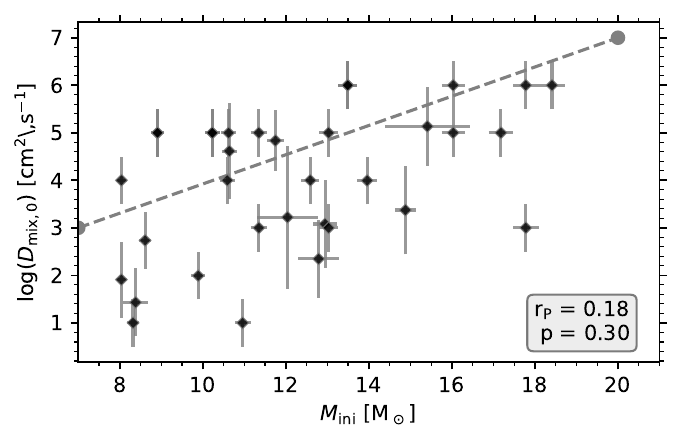}}
    \caption{Envelope mixing strength against initial mass with metallicity fixed to 0.014. Grey circles connected by a dashed line show the mixing coefficient at the base of the envelope in the hydrodynamical simulations of \citet{Varghese2023}. }
    \label{fig:logD_M_fixedZ}
\end{figure}

The Pearson correlation coefficients with the initial mass $M_\mathrm{ini}$ reported in Fig.\,\ref{fig:corner_rotation_mixing_age} are dominated by the two most massive stars. If these two targets are neglected, the correlations with $f_\mathrm{rot}$ and $X'_\mathrm{c}$ become greatly weakened and the latter is no longer significant. Regardless of whether the two outliers are included, there appears to be no strong relation between $M_\mathrm{ini}$ and $\log{D_\mathrm{mix,0}}$ in our sample even though this is predicted from hydrodynamical simulations \citep{Varghese2023} and observations of surface nitrogen abundances \citep{Mombarg2025a}. However, a strong positive correlation occurs if we fix the initial metallicity $Z = 0.014$ and neglect the two outliers. This relation is shown in Fig.\,\ref{fig:logD_M_fixedZ} and compares well with the simulations of \citet{Varghese2023}, which also kept $Z$ fixed to 0.014.

Besides the correlation between $f_\mathrm{ov}$ and $X'_\mathrm{c}$ discussed in Sect.\,\ref{ssec:convective_core_mass}, there is no significant correlations including either of the mixing parameters $f_\mathrm{ov}$ or $\log{D_\mathrm{mix,0}}$. $f_\mathrm{ov}$ is not correlated with $M_\mathrm{ini}$, which is consistent with the findings of \citet{Lechien2026} whose spectroscopy of OB-stars show evidence for $f_\mathrm{ov}$ being independent of $M_\mathrm{ini}$ in this mass range. Notably, $f_\mathrm{ov}$ is also not significantly correlated with $f_\mathrm{rot}$, despite theoretical work suggesting that rotation can reduce the efficiency of convection and core-boundary mixing \citep[e.g.,][]{Tayler1973, Augustson2019, Bessila2025}. Likewise, while \citet{Varghese2024} predicts mixing by internal gravity waves is weakened by rotation, we detect no significant anticorrelation between $\log{D_\mathrm{mix,0}}$ and $f_\mathrm{rot}$. This is in line with previous observational results for B-stars found by \citet{Aerts2014}. 

There are several causes behind this lack of clear relations. First, not all $\beta$\,Cep pulsations are sensitive to the envelope (e.g., the top right panel of Fig.\,\ref{fig:kernels_examples}), which makes it difficult to constrain $\log{D_\mathrm{mix,0}}$ in some $\beta$\,Cep stars. Secondly, the $\beta$\,Cep pulsation class is inherently diverse due to the wide range in mass, age, metallicity, and rotation rate, which complicates the relations between these parameters. Thirdly, the broad parameter space is not covered uniformly or completely by our sample, which features a gap in the mass range between approximately 20 and 25 M$_\sun$ in the mass histogram of Fig.\,\ref{fig:corner_rotation_mixing_age}. Finally, differential rotation is common in $\beta$\,Cep stars (cf. Sect.\,\ref{sec:differential_rotation}), while the optimised $f_\mathrm{rot}$ probe diverse regions of the stars. Therefore, $f_\mathrm{rot}$ may not be representative of the star as a whole, which potentially clouds relations between $f_\mathrm{rot}$ and other free parameters.

\section{Different modelling approaches} \label{app:comparing_modelling}

\subsection{Comparison with \citet{Fritzewski2025}} \label{app:comparing_modelling_Fritzewski}

\begin{figure*}[t]
    \centering
    \includegraphics[width=\linewidth]{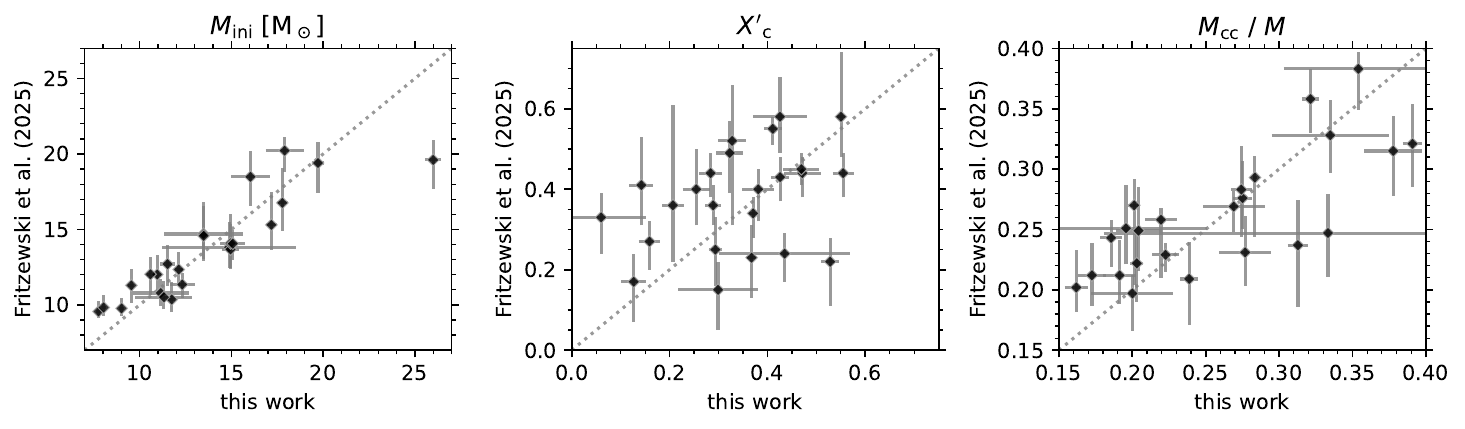}
    \caption{Comparison of the initial mass, normalised central hydrogen mass fraction, and relative convective core mass from our modelling of \nFritzewski stars with the modelling by \citet{Fritzewski2025}. The grey dotted lines indicate where the results from the two modelling procedures agree. }
    \label{fig:comparing_modelling_Fritzewski}
\end{figure*}

Here we test how the addition of more identified frequencies and our more intricate modelling approach improves upon the modelling of \citet{Fritzewski2025}. For the \nFritzewski stars in both samples, we compare three essential parameters in Fig.\,\ref{fig:comparing_modelling_Fritzewski}. Overall, there is a clear agreement in the initial stellar mass $M_\mathrm{ini}$, except at the highest and lowest masses. This is because \citet{Fritzewski2025} used the stellar model grid of \citet{Burssens2023} which only covers $M_\mathrm{ini} \in [9, 21.5]\,M_\sun$, while we found that these stars' masses range from $8\,\mathrm{M_\sun}$ to $30\,\mathrm{M_\sun}$. The differences in the normalised central hydrogen mass fraction $X'_\mathrm{c} = \frac{X_\mathrm{c}}{X_\mathrm{c,ZAMS}}$ are large, demonstrating that this work significantly improved the estimates of the stellar age. This superior $X'_\mathrm{c}$ constraint also results in an improved estimate for the convective core mass due to the strong relation with $X'_\mathrm{c}$ discussed in Sect.\,\ref{ssec:convective_core_mass}.

\subsection{Further evaluating the impact of second-order rotation effects} \label{app:comparing_modelling_GYRE}

As shown in Sect.\,\ref{sec:second_order_rotational_effects}, \texttt{StORM} tends to overestimate the asymmetry of the rotational splitting in $l=1$ multiplets of $\beta$\,Cep stars with a rotation frequency $f_\mathrm{rot}$ greater than $10\%$ of the Keplerian critical rotation frequency $f_\mathrm{crit}$. Consequently, the estimated $f_\mathrm{rot}$ is smaller compared to an extraction of $f_\mathrm{rot}$ from a simplified, `a posteriori' estimate based on the first-order rotation treatment of the \texttt{GYRE} oscillation code. When $f_\mathrm{rot} = 0$, \texttt{StORM} reproduces the frequencies computed with \texttt{GYRE} to within observational errors \citep{Vanlaer2026}. Consequently, we can evaluate the impact of the second-order rotation effects in our forward modelling by comparing the modelling results from these two codes. The most notable second-order effects are the asymmetric rotational splitting and stellar deformation reducing all mode frequencies. 

Here we compare the modelling outcomes from different methodologies in order to answer the following questions: 

\begin{itemize}
    \item Does the reduction of mode frequencies due to stellar deformation and shifts in zonal frequency from mode coupling lead to a better match with the observed zonal frequencies?
    \item Does our consistent inclusion of asymmetric rotational splitting $\Delta f$ in the $\chi^2$ merit function in modelling step 3 worsen the fit of the other observations? 
    \item How does the choice of oscillation code and modelling strategy affect our estimates of the stellar parameters? 
\end{itemize}

We repeated the modelling of our \nModelled stars using the same grid of stellar models, though with the oscillation computations performed with \texttt{GYRE}. The modelling method was identical to the one described in Sect.\,\ref{ssec:modelling_breakdown} with two notable changes. First, zonal mode frequencies $f_0$ are unchanged by rotation to first order. Therefore, the procedure of fixing the age of each evolutionary track using the fixed mode described in Sect.\,\ref{sssec:age_rotation_fixing} was only performed for $f_\mathrm{rot} = 0$. Secondly, the optimal $f_\mathrm{rot}$ is determined from the observed rotational splitting $\Delta f_\mathrm{obs}$ by minimising $\sum_\mathrm{j} (\Delta f_{\mathrm{obs},j} - m_\mathrm{j} (1 - C_{nl,j}) f_\mathrm{rot})^2 / \sigma^2_{\Delta f_{\mathrm{obs},j}}$ with the Ledoux constant $C_{nl}$ \citep{Ledoux1951} computed with \texttt{GYRE}. With this new procedure, the age and $f_\mathrm{rot}$ were again determined for each evolutionary track and only models within $2\sigma$ of the observed position in the HRD were retained. Afterwards, modelling steps 3 to 6 were followed as before. Electronic tables summarising the results using these alternative methods are available at the CDS. 

In another test, we repeated both the analyses with the \texttt{StORM} and \texttt{GYRE} grid with an altered $\chi^2$ merit function in modelling step 3. We left the identified rotational splitting $\Delta f$ out of the merit function, so the statistical parameter estimates are optimised to only the effective temperature $\log{T_\mathrm{eff}}$, luminosity $\log{L}$, and identified zonal mode frequencies $f_0$. 

\begin{figure}
    \resizebox{\hsize}{!}{\includegraphics{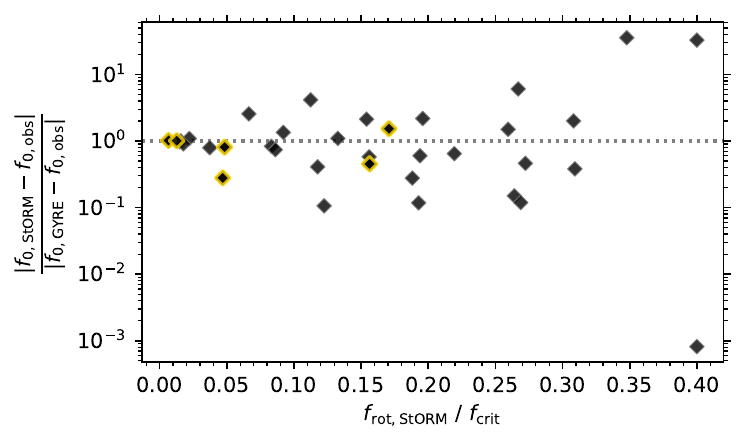}}
    \caption{Ratio of the largest discrepancy in zonal frequency between the observations and models found with \texttt{StORM} over \texttt{GYRE} against the relative rotation frequency. The rotational splitting was not included in the $\chi^2$ merit function in step 3. For stars below the grey dotted line, the modelling using \texttt{StORM} better reproduces the observed zonal modes than the modelling with \texttt{GYRE} and vice versa. The validation stars have a golden outline. }
    \label{fig:comparing_fZonal_GYRE_StORM}
\end{figure}

We first examine the impact of second-order rotation effects on the modelling of zonal modes by using the results found when neglecting the contributions of $\Delta f$ in modelling step 3. Figure\,\ref{fig:comparing_fZonal_GYRE_StORM} compares the largest discrepancy between the observed $f_0$ and the those in the model found with \texttt{StORM} and \texttt{GYRE}. Most stars lie well below the grey dotted line, which indicates that the \texttt{StORM} model outperforms the \texttt{GYRE} model. Therefore, we conclude that the treatment of rotationally induced stellar deformation and mode coupling in \texttt{StORM} improves the fitting of $f_0$. 

\begin{figure}
    \resizebox{\hsize}{!}{\includegraphics{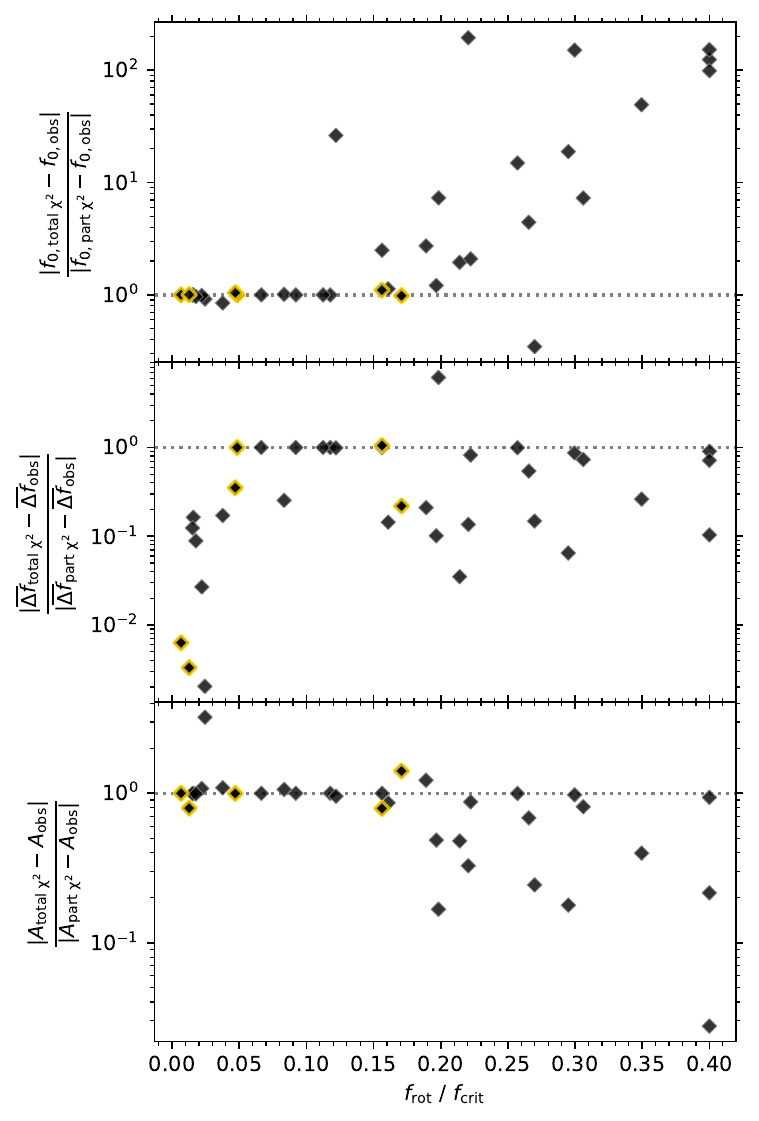}}
    \caption{Ratio of the largest discrepancy in zonal mode frequencies (top), mean splitting (middle), and asymmetry (bottom) between observations and models found when including over excluding the rotational splitting in the $\chi^2$ merit function in step 3 against the relative rotation frequency. For stars below the grey dotted line, the modelling with rotational splitting in $\chi^2$ outperforms that without and vice versa. The validation stars have a golden outline. }
    \label{fig:comparing_all_splittingChi2}
\end{figure}

\begin{figure*}[t]
    \centering
    \includegraphics[width=\linewidth]{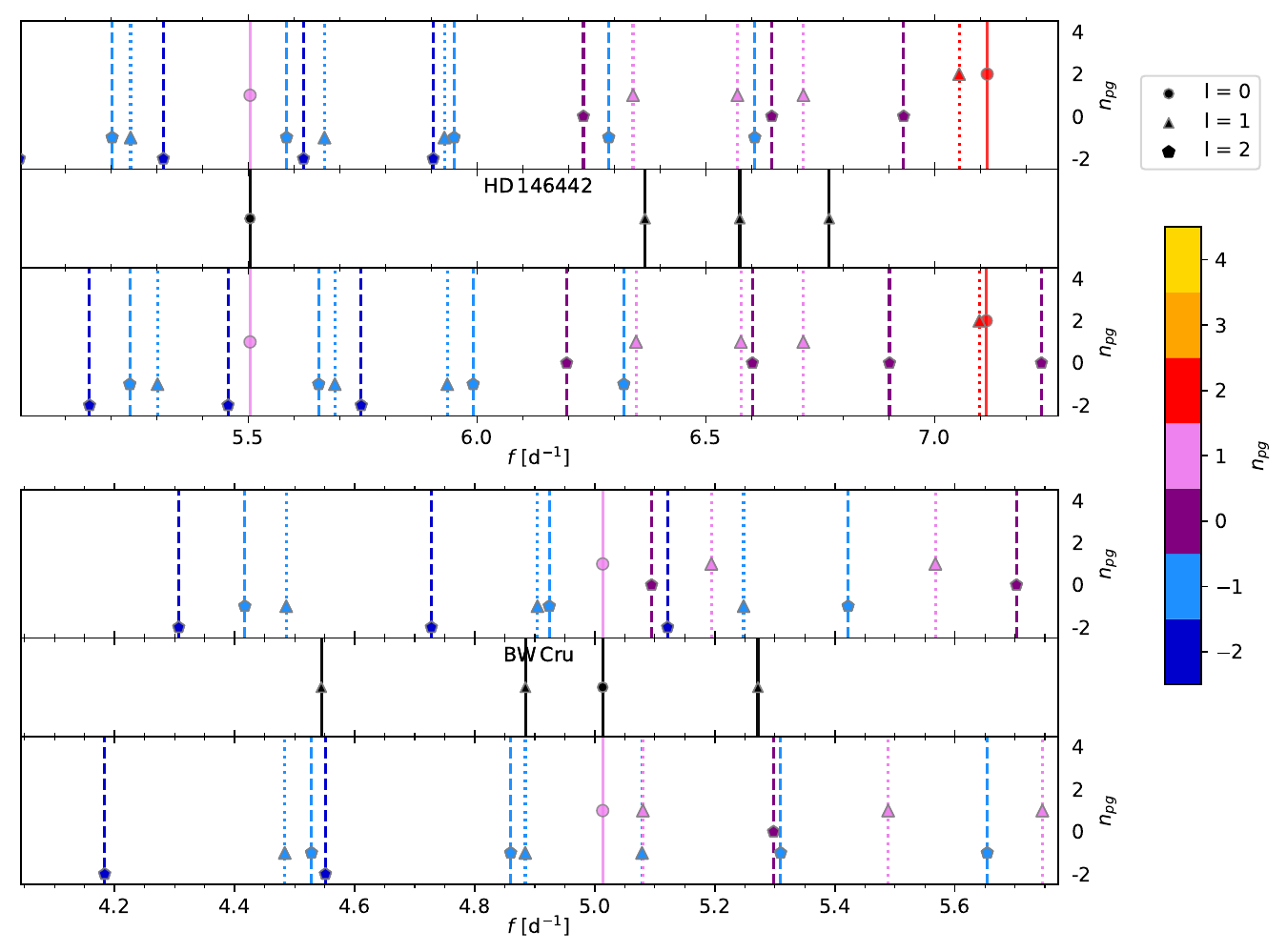}
    \caption{Idem Fig.\,\ref{fig:frequency_fits_validationStars}, comparing two modelling approaches to observations for two example stars. For each star, three panels show the frequencies in the models found when including (top) and excluding (bottom) rotational splitting in the $\chi^2$ merit function in modelling step 3 (top) and the observations (middle) -- neglecting amplitude and unidentified frequencies for clarity. }
    \label{fig:comparing_frequency_fits_splittingChi2}
\end{figure*}

Next, Fig.\,\ref{fig:comparing_all_splittingChi2} compares the model quality with and without the contribution of $\Delta f$ to $\chi^2$ using the \texttt{StORM} grid. On one hand, the observed $f_0$ are generally better reproduced when $\Delta f$ is not optimised alongside $f_0$ in $\beta$\,Cep stars rotating more rapidly than $\sim20\%\,f_\mathrm{crit}$. This indicates that beyond this rotation threshold, the difficulty in matching the observed rotational splitting may throw off the fitting of zonal frequencies. Figure\,\ref{fig:comparing_frequency_fits_splittingChi2} shows two such stars, where one's model is weakly affected and the other's strongly affected by the change step 3. On the other hand, the observed asymmetry and mean rotational splitting $\overline{\Delta f}$ are naturally better reproduced when $\Delta f$ is included. Notably, $\overline{\Delta f}$ is much better reproduced when including second-order effects at small $f_\mathrm{rot} / f_\mathrm{crit}$. These differences are exaggerated at small $f_\mathrm{rot} / f_\mathrm{crit}$ as $\overline{\Delta f}$ is reproduced precisely in both approaches, which makes the ratio between the two models volatile. Nevertheless, this demonstrates that optimising asymmetric rotational splitting from second-order effects is worthwhile even  for slow rotators, as also argued by \citet{Briquet2007}. 

In summary, the inclusion of $\Delta f$ in $\chi^2$ can adversely affect the fitting of $f_0$ when $f_\mathrm{rot} > 20\%\,f_\mathrm{crit}$ and potentially affect stellar parameters such as mass and age. However, it also leads to better reproduction of non-zonal frequencies and thus produces a superior $f_\mathrm{rot}$ estimate at all rotation rates. Consequently, whether $\Delta f$ should be optimised consistently depends the rotation regime as well as on which stellar parameters one prioritises. 

\begin{figure*}[t]
    \centering
    \includegraphics[width=\linewidth]{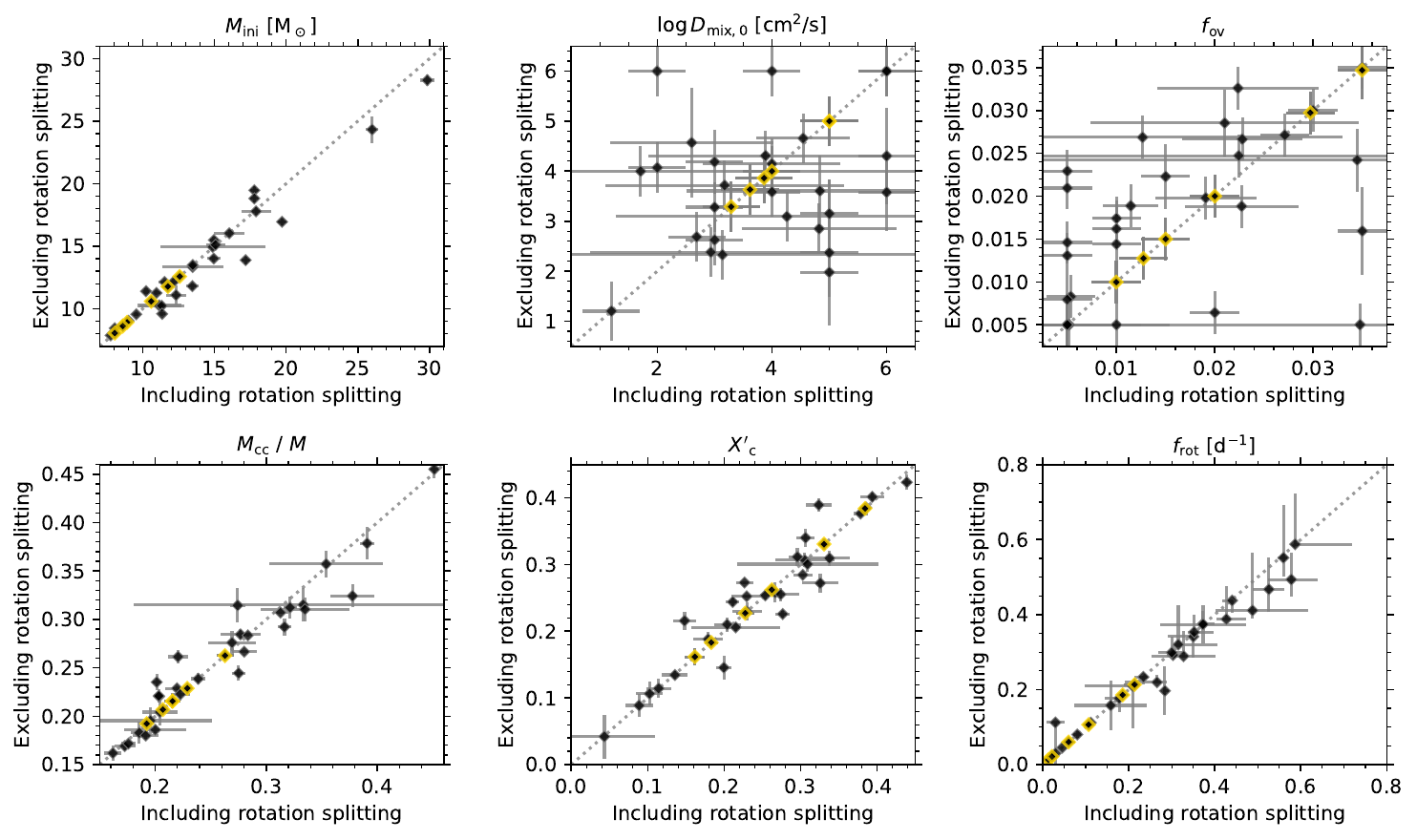}
    \caption{Comparison of the initial mass, envelope mixing strength, core overshoot parameter, relative convective core mass, normalised central hydrogen mass fraction, and rotation frequency found when including and excluding the rotational splitting in $\chi^2$ in modelling step 3. Grey dotted lines indicate where the modelling results agree. The validation stars have a golden outline. }
    \label{fig:comparing_modelling_splittingChi2}
\end{figure*}

To test how these different modelling approaches affect the derived stellar parameters, Fig.\,\ref{fig:comparing_modelling_splittingChi2} compares six parameters obtained when including or excluding $\Delta f$ in $\chi^2$. By and large, these modelling approaches produce similar results as the initial mass $M_\mathrm{ini}$ and the normalised central hydrogen mass fraction $X'_\mathrm{c} = \frac{X_\mathrm{c}}{X_\mathrm{c,ZAMS}}$ as a proxy of age agree for most stars. Subsequently, $M_\mathrm{cc} / M$ is also in agreement. Nonetheless, there are a handful of stars for which the two modelling procedures differ significantly as a local minimum in $\chi^2$ becomes the global minimum. 

\begin{figure*}[t]
    \centering
    \includegraphics[width=\linewidth]{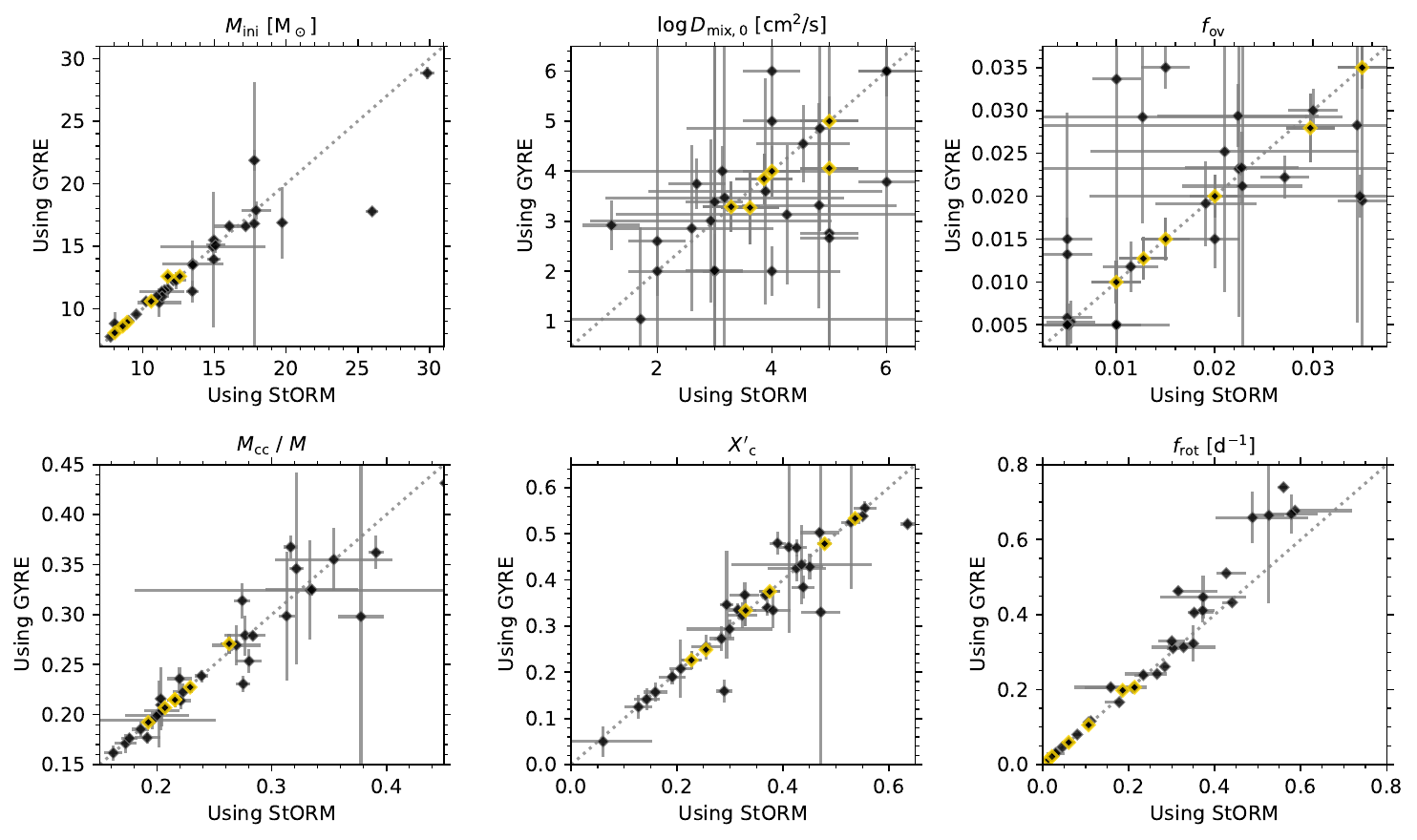}
    \caption{Comparison of the initial mass, envelope mixing strength, core overshoot parameter, relative convective core mass, normalised central hydrogen mass fraction, and rotation frequency found from \texttt{StORM} and \texttt{GYRE}. Grey dotted lines indicate where the modelling results agree. The validation stars have a golden outline. }
    \label{fig:comparing_modelling_GYRE_StORM}
\end{figure*}

For completeness, we also compare the modelling results obtained with \texttt{StORM} and \texttt{GYRE} in Fig.\,\ref{fig:comparing_modelling_GYRE_StORM} with $\Delta f$ considered in $\chi^2$. Again, most parameters are broadly in agreement, although $X'_\mathrm{c}$ tends to get underestimated when using \texttt{GYRE} at higher rotation rates because the reduction of mode frequencies due to stellar deformation is neglected. This produces significant differences in the estimates of the other parameters. Notably, these differences are larger than the scatter in Fig.\,\ref{fig:comparing_modelling_splittingChi2}, which suggests that the choice of oscillation code has a greater impact than the modelling methodology. As also discussed in Sect.\,\ref{sec:second_order_rotational_effects}, the optimal $f_\mathrm{rot}$ is usually greater when using \texttt{GYRE} than when using \texttt{StORM}, especially at higher $f_\mathrm{rot}$.

\section{Sensitivity kernels and their evolution} \label{app:sensitivity_kernels}

\begin{figure*}[h]
    \sidecaption
    \includegraphics[width=11cm]{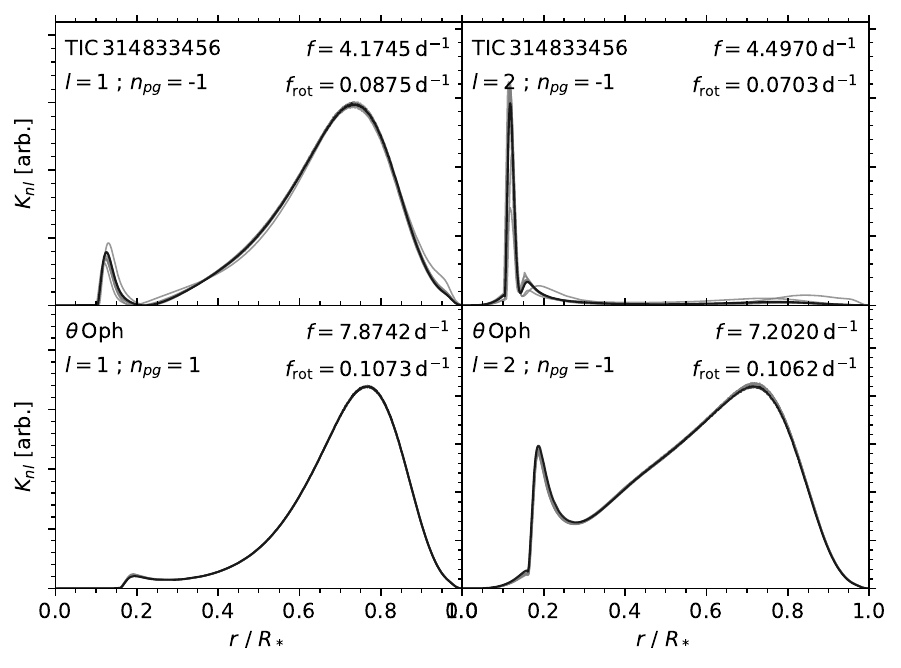}
    \caption{Sensitivity kernels of the identified rotationally split multiplets of two stars in black. Thin grey lines show the sensitivity kernels in models with one model parameter increased and decreased by $1\sigma$. In each panel, we include the name of the star and the multiplet's degree, radial order, zonal frequency, and extracted rotation frequency. }
    \label{fig:kernels_examples}
\end{figure*}

The low-radial order pulsations in $\beta$\,Cep stars typically have a broad rotational sensitivity kernel $K_{nl}$. Figure\,\ref{fig:kernels_examples} displays $K_{nl}$, calculated using equation (3.356) in \citet{Aerts2010}, of the identified rotationally split multiplets in two stars. TIC\,314833456 (top panel) has one multiplet sensitive to a thin region near the core and another sensitive to a broad part of the envelope, hence its core-to-envelope rotation ratio can be well constrained. Meanwhile, $\theta$\,Oph (bottom panel) has two multiplets that are both sensitive to similar broad regions of the envelope. Consequently, these overlapping kernels average the rotation frequency over the same values, dampening the difference between the two measured rotation frequencies. Therefore, our constraint on the differential rotation in $\theta\,$Oph shown in Fig.\,\ref{fig:diffrot_log_atMode_against_Xc} and reported in Table\,\ref{tab:differential_rotation} is a lower limit on the true difference in rotation rate throughout the star.

To examine how the sensitivity kernels depend on the model parameters, we computed \texttt{MESA-StORM} models for $\theta$\,Oph and TIC\,314833456 at the optimal model parameters $Z$, $M_\mathrm{ini}$, $\log{D}_\mathrm{mix,0}$, $f_\mathrm{ov}$, and $X_\mathrm{c}$ from modelling step 3, except with one parameter increased or decreased by $1\sigma$. Since the model parameters are strongly correlated, this is somewhat unrealistic, yet it serves to show the shapes $K_{nl}$ can take on. The resulting sensitivity kernels are shown as thin grey lines in Fig.\,\ref{fig:kernels_examples}. Thanks to $\theta$\,Oph's high-precision parameters and relatively young age, its sensitivity kernels hardly vary. TIC\,314833456 is more evolved and has larger parameter errors, resulting in the kernel of the $l=1$ mode varying somewhat more. This emphasises that the positions on the x-axis shown in the left panel of Fig.\,\ref{fig:diffrot_log_atMode_against_Xc} are only approximate, yet sufficiently stable to draw conclusions. 

The left panel of Fig.\,\ref{fig:diffrot_log_atMode_against_Xc} shows that six stars have several multiplets which probe the upper envelope. The right panel of Fig.\,\ref{fig:diffrot_log_atMode_against_Xc} shows these six have $X'_\mathrm{c} > 0.42$, making them among the least evolved stars of this subsample of \nMultiple stars. Meanwhile, ten out of eleven stars with both their near-core region and outer envelope probed are more evolved, with $X'_\mathrm{c} < 0.38 $. This is in excellent agreement with computations of mode excitation in a 9\,M$_\sun$ stellar model by \citet{Rehm2024}, which indicate that several p-modes, which mostly probe the envelope, are only excited while the star is young. Once the star is more evolved, it excites more and higher-order g-modes, which are most sensitive to the near-core layers. 

This age-dependent mode excitation complicates constraining the rotation profiles of \BCs at different ages. Indeed, we did not measure the envelope rotation gradient in evolved \BC star and mostly measured the core-to-envelope rotation ratio in relatively young $\beta$\,Cep pulsators. Nevertheless, such detections are worth pursuing. Moreover, this relation between age and the sensitivity kernels may inform the targetting decisions of future observational studies that seek to study angular momentum transport.

\section{Statistical test of differential rotation} \label{app:statistical_test}

In this section, we describe the test used to determine if the measurements of rotation from different multiplets are significantly different. This is complicated by the unidirectional systematic error on each measurement of rotation described in Sect.\,\ref{sec:second_order_rotational_effects}. The systematic error is not known precisely, but only constrained by a bound. 

We write each rotation measurement $Q_i$ as 
\begin{equation}
    Q_i = q_i + \sigma_i - b_i
\end{equation}
with $q_i$ the true value, $\sigma_i$ the uncertainty on the measurement, which is assumed to be Gaussian, and $b_i$ the systematic error. This systematic error is not known, but it is bounded by a limit $B_i$ such that $|b_i| < |B_i|$. Since the systematic can be either positive or negative, we may rewrite this condition as $b_i \in [L_i, U_i]$, with $L_i$ and $U_i$ the lower and upper bounds on the possible values of the systematic offset. In practise, one of $L_i$ or $U_i$ will be zero because the systematic error is unidirectional. 

The null hypothesis $H_0$ of our statistical test comparing two measurements of rotation $Q_1$ and $Q_2$ is $q_1 = q_2 =\theta$. As the test statistic, we use the likelihood ratio $\lambda = \frac{\max(\mathcal{L_{H_0}}(b_1, b_2))}{\max{(\mathcal{L}(b_1, b_2))}}$ with the likelihood $\mathcal{L}(b_1, b_2) = \exp{\left( - \frac{(Q_1 + b_1 - q_1)^2}{2 \sigma_1^2} - \frac{(Q_2 + b_2 - q_2)^2}{2 \sigma^2} \right)}$. Consequently, $\mathcal{L}_{H_0} = \exp{\left( - \frac{(Q_1 + b_1 - \theta)^2}{2 \sigma_1^2} - \frac{(Q_2 + b_2 - \theta)^2}{2 \sigma^2} \right)}$. $\max{(\mathcal{L})}_{H_0}$ is the highest likelihood that can be attained without $b_1$ or $b_2$ leaving their constrained ranges and assuming that $q_1 = q_2 = \theta$. Since $q_1$ and $q_2$ vary independently in $\lambda$'s denominator, they can take on the values $Q_1 + b_1$ and $Q_2 + b_2$, respectively, so $\max{(\mathcal{L})} = 1$. 

For given $(b_1, b_2)$, the likelihood in the numerator is maximal if 
\begin{equation*}
    \theta = \frac{(Q_1 + b_1)^2\sigma_2^2 + (Q_2 + b_2)^2 \sigma_1^2}{\sigma_1 ^2 + \sigma_2^2} \ ,
\end{equation*}
in which case the log-likelihood $l$ is 
\begin{equation*}
    l = \ln{\mathcal{L}_{H_0}} = -\frac{1}{2} \frac{\left[ (Q_1 + b_1) - (Q_2 + b_2) \right]^2}{\sigma_1 ^2 + \sigma_2^2} \ . 
\end{equation*}
As such, the $\mathcal{L}_{H_0}$ is maximised by minimising $|(Q_1 + b_1) - (Q_2 + b_2)|$. Defining $d = Q_1 - Q_2$ and $i = b_1 - b_2$, we thus seek $\delta = \min{|d + i|}$ while constrained by $i \in [L_1 - U_2, U_1 - L_2]$. 

Once the minimal $\delta$ is found, the likelihood ratio is computed as  
\begin{equation}
    \lambda = \exp{\left( -\frac{1}{2} \frac{\delta^2}{\sigma_1^2 + \sigma_2^2} \right)} \ .
\end{equation} 
We point out that $\lambda$ is not a p-value. Instead, a result as $\lambda = 0.2$ can be interpreted as the given $Q_1$ and $Q_2$ being five times less likely to get measured if $q_1 = q_2$ than if the true underlying values $q_1$ and $q_2$ are different. In other words, the measurements are five times less likely to be taken in a uniformly rotating star as compared to a differentially rotating star. Should the systematic errors be zero, $-2 \ln{\lambda}$ would follow a $\chi^2$ distribution and $\lambda < 0.147$ would correspond to a p-value $p < 0.05$. Therefore, we use $\lambda < 0.1$ as a conservative threshold to consider a star a differential rotator.

\section{Inequalitities of \citet{Reese2015}} \label{app:Reese_test}

\citet{Reese2015} derived a number of inequalities that the rotational splittings should obey if a star's rotation profile decreases monotonically with radius. For our purposes, only the inequality derived in their Sect.\,2.1.1 is applicable since the core rotation gradient is unconstrained, most stars only have two rotationally split multiplets, and the modest levels of differential rotation found in Sect.\,\ref{sec:differential_rotation} ensure that the Rayleigh criterion is maintained. The relevant inequality states that 
\begin{equation} \label{eq:Reese_inequality}
    a s'_i \leq s'_j \leq b s'_i\, ,
\end{equation}
with $s'_i = \frac{s_i}{1 - C_{i}}$, where $s_i$ is the rotational splitting of multiplet $i$ corrected for the Ledoux constant $C_i$. As such, $s_i$ is effectively an estimate of the rotation rate. The coefficients $a$ and $b$ are the minimum and maximum ratios of the integrated sensitivity kernels, specifically
\begin{align}
    I_i(r) &= \int_0^r K_{i} \, dr \\
    a &= \min\left( \frac{I_i(r)}{I_j(r)} \right) \\
    b &= \max\left( \frac{I_i(r)}{I_j(r)} \right)
\end{align}
, wherein $K_i$ is the rotation sensitivity kernel of the multiplet. 
As explained in \citet{Reese2015}, this inequality is only applicable to pairs of multiplets of equal degree $l$\footnote{Except for pairings of $l=1$ and $l=3$ modes, but we do not consider $l > 2$ modes in this work.} and only if their integrated kernels $I_i$ do not cross. In the \nMultiple stars in our sample with more than one rotationally split multiplet, there are 30 pairs of multiplets of equal $l$. The integrated kernels cross in 21 of those 30 pairs, leaving only nine pairings in five stars to test the inequality. Of these nine, all but one pairing of $l=2$ modes in HD\,192575 obey the inequality \ref{eq:Reese_inequality}. 

If a pair of rotationally split multiplets breaks this inequality, it either indicates the star's rotation profile does not decrease monotonically or the stellar model from which the rotational sensitivity kernels are derived is inaccurate. HD\,192575 was already known to have a non-monotonic rotating profile \citep{Vanlaer2025a}, which we confirmed in Sect.\,\ref{sec:differential_rotation}, so we expected it to break inequality \ref{eq:Reese_inequality}. Due to the limited constraining power of inequality \ref{eq:Reese_inequality} when only two or three multiplets are identified, the rotation profile of the remaining four stars are not guaranteed to be monotonically decreasing. Indeed, two other pairings of multiplets in HD\,192575 obey inequality \ref{eq:Reese_inequality}, despite its non-monotonic rotation profile.

\section{Electronic data, code, and figures}

This paper is accompanied by a number of electronic tables containing all the observational constraints as well as all the modelling results required to reproduce the figures presented in the main text. Snippets of these tables are shown in this Appendix. Further instructions on how to access these electronic tables, electronic figures, and our forward modelling code are given in the Data Availability section above.  

\begin{table*}[t]
    \centering
    \caption{Observational input used in the modelling of five stars examined in this study, including mode identities. For brevity, we omitted the uncertainties and show only one identified oscillation mode. }
    \resizebox{\linewidth}{!}{
    \begin{tabular}{lllllllllllllllllll}
        \hline\hline
        TIC ID & Name & \textit{Gaia} DR3 ID & RA & dec. & $V$ & success & source $T_\mathrm{eff}$ & $\log{T_\mathrm{eff}}$ & $\log{L}$ & $f_\mathrm{rot,surface}\sin{i}$ & $N_\mathrm{i}$ & $N_\mathrm{f}$ & $i_1$ & $n_\mathrm{pg,1}$ & $l_1$ & $m_1$ & $f_1$ & $a_1$ \\ 
         &  &  & [\degr] & [\degr] & [mag] &  &  & [K] & [L$_\sun$] & [d$^{-1}$] &  &  &  &  &  &  & [d$^{-1}$] &  \\ 
        \hline
        13332837 & HD$\,$229085 & 2061190956100233088 & 305.39636 & 38.61325 & 9.8 & True & esphs & 4.366 & 4.172 & 0.48 & 2 & 5 & 1 & 2 & 1 & -1 & 8.638539 & 0.001555 \\
        14085632 & TIC$\,$14085632 & 2057943789022548096 & 305.72188 & 37.11278 & 11.0 & True & esphs & 4.533 & 4.938 & 0.38 & 2 & 7 & 1 & 1 & 1 & -1 & 4.521608 & 0.003241 \\
        15166556 & HD$\,$146442 & 5990434159009246848 & 244.59709 & -45.84034 & 9.11 & True & esphs & 4.362 & 3.727 & 0.76 & 2 & 4 & 1 & 1 & 1 & -1 & 6.366871 & 0.002721 \\
        18827544 & TIC$\,$18827544 & 5941164183970835200 & 247.97767 & -48.42875 & 12.28 & False & esphs & 4.398 & 3.939 & 0.38 & 2 & 4 & 1 & \dots & 1 & -1 & 4.39127 & 0.000732 \\
        34590771 & $\beta\,$CMa & -1 & 95.67494 & -17.95592 & 1.97 & True & Mazumdar2006 & 4.4 & 4.45 & \dots & 2 & 3 & 1 & 1 & 0 & 0 & 3.9995 & 2.6 \\
        \dots \\
        \hline
    \end{tabular}}
    \tablefoot{$N_\mathrm{i}$ is the number of identified radial and zonal modes and $N_\mathrm{f}$ is the total number of identified modes. The column `success' indicates if we found a satisfactory model. The columns suffixed with $_1$ describe the first identified pulsation mode. The complete table, including uncertainties, all identified frequencies, and lists of all detected frequencies including unidentified ones, is available at the CDS with additional documentation. }
    % \tablebib{(1) \citet{Mazumdar2006}; (2) \citet{Briquet2007}; (3) \citet{Aerts2003b}; (4) \citet{DeRidder2004}; (5) \citet{Burssens2023}.}
    \label{tab:observational_input}
\end{table*}

\begin{table*}[t]
    \centering
    \caption{Modelling outcomes of five stars successfully modelled in this study using \texttt{StORM} and including rotational splitting in the $\chi^2$ merit function in step 3. For simplicity, we omitted the uncertainties in this excerpt. }
    \resizebox{\linewidth}{!}{
    \begin{tabular}{lllllllllllllllllllll}
        \hline\hline
        TIC ID & Name & $\chi^2$ & $\max{|f_{0,\mathrm{obs}} - f_{0,\mathrm{model}}|}$ & $Z$ & $M_\mathrm{ini}$ & $\log{D_\mathrm{mix,0}}$ & $f_\mathrm{ov}$ & $X_\mathrm{c}$ & $X'_\mathrm{c}$ & age & $f_\mathrm{rot}$ & $f_\mathrm{rot}/f_\mathrm{crit}$ & $v_\mathrm{rot}$ & $\log{T_\mathrm{eff}}$ & $\log{L}$ & $\log{g}$ & $\log{R}$ & $M$ & $M_\mathrm{cc}$ & $M_\mathrm{cc}/M$ \\
         &  &  & [d$^{-1}$] &  & [M$_\sun$] & [cm$^2$\,s] &  &  &  & [Myr] & [d$^{-1}$] &  & [km$\,$s$^{-1}$] & [K] & [L$_\sun$] & [cm\,s$^2$] & [R$_\sun$] & [M$_\sun$] & [M$_\sun$] &  \\
        \hline
        13332837 & HD$\,$229085 & 12199.1 & 0.0001256 & 0.011 & 10.23 & 3.0 & 0.01 & 0.324 & 0.451 & 16.27 & 0.4268 & 0.2141 & 126.0 & 4.376 & 3.992 & 3.915 & 0.766 & 10.21 & 2.25 & 0.221 \\
        14085632 & TIC$\,$14085632 & 314.3 & 0.0023416 & 0.0116 & 17.9 & 6.0 & 0.0224 & 0.305 & 0.426 & 8.66 & 0.1118 & 0.0834 & 52.0 & 4.478 & 4.795 & 3.762 & 0.964 & 17.85 & 6.32 & 0.354 \\
        15166556 & HD$\,$146442 & 1570.0 & 0.0051135 & 0.011 & 7.76 & 5.0 & 0.005 & 0.204 & 0.284 & 35.53 & 0.372 & 0.2223 & 113.5 & 4.297 & 3.705 & 3.766 & 0.781 & 7.75 & 1.34 & 0.173 \\
        34590771 & $\beta\,$CMa & 0.0 & 0.0009837 & 0.0133 & 12.58 & 3.3 & 0.0127 & 0.162 & 0.227 & 14.66 & 0.0602 & 0.0484 & 26.6 & 4.38 & 4.355 & 3.653 & 0.941 & 12.55 & 2.7 & 0.215 \\
        42940133 & HD$\,$228101 & 184.4 & 0.0013233 & 0.014 & 11.75 & 4.8 & 0.021 & 0.394 & 0.555 & 12.56 & 0.234 & 0.1177 & 73.2 & 4.411 & 4.183 & 3.926 & 0.791 & 11.75 & 3.25 & 0.277 \\
        \dots \\
        \hline
    \end{tabular}}
    \tablefoot{The full table, including uncertainties, is available at the CDS with additional documentation. Similar tables using the alternative modelling methods described in Appendix\,\ref{app:comparing_modelling_GYRE} are also provided. }
    \label{tab:statistical_models}
\end{table*}

\begin{table*}[t]
    \centering
    \caption{Observed and modelled rotational splitting in each identified multiplet of five stars in our sample. }
    \resizebox{\linewidth}{!}{
    \begin{tabular}{lllllllllllllllllllllll}
        \hline\hline
        TIC ID & Name & $n_\mathrm{pg}$ & $l$ & $f_{0,\mathrm{obs}}$ & $f_{0,\mathrm{model}}$ & $\Delta f_{1,\mathrm{obs}}$ & $\Delta f_{-1,\mathrm{obs}}$ & $\Delta f_{2,\mathrm{obs}}$ & $\Delta f_{-2,\mathrm{obs}}$ & $A_{1,\mathrm{obs}}$ & $A_{2,\mathrm{obs}}$ & $\Delta f_{1,\mathrm{model}}$ & $\Delta f_{-1,\mathrm{model}}$ & $\Delta f_{2,\mathrm{model}}$ & $\Delta f_{-2,\mathrm{model}}$ & $A_{1,\mathrm{model}}$ & $A_{2,\mathrm{model}}$ & $f_\mathrm{rot}$ & $\sigma_{f_\mathrm{rot}}$ & $b_{f_\mathrm{rot}}$ & $f_\mathrm{rot,\texttt{GYRE}}$ & $C_{nl}$ \\
         &  &  &  & [d$^{-1}$] & [d$^{-1}$] & [d$^{-1}$] & [d$^{-1}$] & [d$^{-1}$] & [d$^{-1}$] &  &  & [d$^{-1}$] & [d$^{-1}$] & [d$^{-1}$] & [d$^{-1}$] &  &  & [d$^{-1}$] & [d$^{-1}$] & [d$^{-1}$] & [d$^{-1}$] & \\
        \hline
        13332837 & HD$\,$229085 & 2 & 1 & 9.13184 & 9.13184 & 0.59119 & 0.4933 & \dots & \dots & -0.09026 & \dots & 0.47734 & 0.61843 & \dots & \dots & 0.12876 & \dots & 0.4501 & 0.1168 & 0.1761 & 0.559 & 0.03002 \\
        13332837 & HD$\,$229085 & 1 & 1 & 6.87714 & 6.87727 & \dots & 0.3861 & \dots & \dots & \dots & \dots & 0.2668 & 0.38614 & \dots & \dots & 0.18277 & \dots & 0.3421 & 0.0019 & -0.0 & 0.4021 & 0.03974 \\
        14085632 & TIC$\,$14085632 & 1 & 1 & 4.63305 & 4.63305 & 0.1386 & 0.11144 & \dots & \dots & -0.10864 & \dots & 0.10805 & 0.13196 & \dots & \dots & 0.09963 & \dots & 0.1275 & 0.0014 & 0.0276 & 0.1328 & 0.05872 \\
        14085632 & TIC$\,$14085632 & -1 & 2 & 4.21651 & 4.21417 & 0.09717 & 0.07341 & \dots & 0.15961 & -0.13931 & \dots & 0.07814 & 0.08072 & 0.15341 & 0.16473 & 0.01627 & 0.03558 & 0.1035 & 0.0033 & 0.0056 & 0.1117 & 0.23358 \\
        15166556 & HD$\,$146442 & 1 & 1 & 6.57384 & 6.56873 & 0.19547 & 0.20697 & \dots & \dots & 0.02857 & \dots & 0.14411 & 0.22807 & \dots & \dots & 0.22559 & \dots & 0.372 & 0.01 & 0.028 & 0.409 & 0.50802 \\
        15166556 & HD$\,$146442 & 1 & 0 & 5.50359 & 5.50359 & \dots & \dots & \dots & \dots & \dots & \dots & \dots & \dots & \dots & \dots & \dots & \dots & \dots & \dots & \dots & \dots & \dots \\
        34590771 & $\beta\,$CMa & 1 & 0 & 3.9995 & 3.9995 & \dots & \dots & \dots & \dots & \dots & \dots & \dots & \dots & \dots & \dots & \dots & \dots & \dots & \dots & \dots & \dots & \dots \\
        34590771 & $\beta\,$CMa & -2 & 2 & 3.8828 & 3.88182 & \dots & \dots & 0.0965 & \dots & \dots & \dots & 0.04866 & 0.04948 & 0.09649 & 0.09987 & 0.0084 & 0.01717 & 0.0595 & 0.0013 & 0.0 & 0.0602 & 0.19842 \\
        42940133 & HD$\,$228101 & 0 & 2 & 7.1674 & 7.16607 & 0.22259 & 0.24275 & \dots & 0.43567 & 0.04331 & \dots & 0.21416 & 0.2214 & 0.42089 & 0.44987 & 0.01662 & 0.03328 & 0.234 & 0.0114 & -0.0021 & 0.235 & 0.0661 \\
        42940133 & HD$\,$228101 & 1 & 0 & 6.30245 & 6.30245 & \dots & \dots & \dots & \dots & \dots & \dots & \dots & \dots & \dots & \dots & \dots & \dots & \dots & \dots & \dots & \dots & \dots \\
        \dots \\
        \hline
    \end{tabular}}
    \tablefoot{Each row in this table represents one rotationally split multiplet or radial mode, meaning there are several rows per star. We included the statistical rotation frequency estimate from our modelling step 3, its uncertainty and bias, and the rotation frequency found with the simplified `a posteriori' step from the Ledoux constant estimated with \texttt{GYRE}. In the subscripts of zonal frequencies $f_{0,\mathrm{source}}$, rotational splitting $\Delta f_{m,\mathrm{source}}$ and asymmetry $A_{|m|,\mathrm{source}}$, `$m$' indicates the azimuthal order and `source' whether the value comes from the observations or the best model. The full table is available at the CDS with additional documentation. }
    \label{tab:rotational_splitting}
\end{table*}

\begin{table*}[t]
    \centering
    \caption{The constraints on rotation in five stars in our sample. We only show the rotation and kernel positions from one rotationally split multiplet in this excerpt for brevity. }
    \resizebox{\linewidth}{!}{
    \begin{tabular}{llllllllllllllllllllllllll}
        \hline\hline
        TIC ID & Name & $f_\mathrm{rot,surface} \sin{i}$ & $\sigma_{f_\mathrm{rot,surface} \sin{i}}$ & $f_\mathrm{rot}$ & $\sigma_{f_\mathrm{rot}}$ & $b_{f_\mathrm{rot}}$ & $X'_\mathrm{c}$ & $\sigma_{X'_\mathrm{c}}$ & $N_\mathrm{r}$ & $f_{0,1}$ & $n_1$ & $l_1$ & $f_\mathrm{rot,1}$ & $\sigma_{f_\mathrm{rot,1}}$ & $b_{f_\mathrm{rot,1,+}}$ & $(r/R_*)_\mathrm{mode,1}$ & $(r/R_*)_\mathrm{mean,1}$ & $(r/R_*)_\mathrm{median,1}$ & $\mathbf{(r/R_*)_\mathrm{FWHM,1}}$ & $\mathbf{(r/R_*)_\mathrm{P25,1}}$ & $\mathbf{(r/R_*)_\mathrm{P75,1}}$ \\
         &  & [d$^{-1}$] & [d$^{-1}$] & [d$^{-1}$] & [d$^{-1}$] & [d$^{-1}$] &  &  &  & [d$^{-1}$] &  &  & [d$^{-1}$] & [d$^{-1}$] & [d$^{-1}$] &  &  &  &  &  &  & \\
        \hline
        13332837 & HD$\,$229085 & 0.48 & 0.15 & 0.4268 & 0.0229 & 0.0232 & 0.451 & 0.026 & 2 & 9.13184 & 2 & 1 & 0.4501 & 0.1168 & 0.1761 & 0.882 & 0.738 & 0.837 & 0.125 & 0.584 & 0.904 \\
        14085632 & TIC$\,$14085632 & 0.38 & 0.1 & 0.1118 & 0.0026 & -0.0025 & 0.426 & 0.055 & 2 & 4.63305 & 1 & 1 & 0.1275 & 0.0014 & 0.0276 & 0.744 & 0.431 & 0.437 & 0.255 & 0.606 & 0.782 \\
        15166556 & HD$\,$146442 & 0.76 & 0.23 & 0.372 & 0.01 & 0.0167 & 0.284 & 0.023 & 1 & 6.57384 & 1 & 1 & 0.372 & 0.01 & 0.028 & 0.131 & 0.451 & 0.196 & 0.051 & 0.135 & 0.805 \\
        34590771 & $\beta\,$CMa & \dots & \dots & 0.0602 & 0.0013 & 0.0115 & 0.227 & 0.02 & 1 & 3.8828 & -2 & 2 & 0.0595 & 0.0013 & 0.0 & 0.123 & 0.406 & 0.384 & 0.017 & 0.149 & 0.684 \\
        42940133 & HD$\,$228101 & 0.46 & 0.12 & 0.234 & 0.0114 & -0.0103 & 0.555 & 0.021 & 1 & 7.1674 & 0 & 2 & 0.234 & 0.0114 & -0.0021 & 0.205 & 0.405 & 0.265 & 0.088 & 0.202 & 0.694 \\
        \dots \\
        \hline
    \end{tabular}}
    \tablefoot{$N_r$ is the number of identified rotationally split multiplets. The columns $(r/R_*)_\mathrm{X}$ are the relative radius of the position of the maximum, mean, median, full-width at half maximum, and 25\% and 75\% percentiles of the sensitivity kernel. The columns suffixed with $_1$ describe the first identified multiplet. The full table, including all identified rotationally split multiplets, is available at the CDS with additional documentation. }
    \label{tab:differential_rotation}
\end{table*}

\end{appendix}

\end{document}